%% file: CMS-ttbar-Lambda.tex
\pdfoutput=1
\documentclass{SciPost}

\hypersetup{
    colorlinks,
    linkcolor={red!50!black},
    citecolor={blue!50!black},
    urlcolor={blue!80!black}
}

\usepackage[bitstream-charter]{mathdesign}
\DeclareSymbolFont{usualmathcal}{OMS}{cmsy}{m}{n}
\DeclareSymbolFontAlphabet{\mathcal}{usualmathcal}

\fancypagestyle{SPstyle}{
\fancyhf{}
\lhead{\colorbox{scipostblue}{\bf \color{white} ~SciPost Physics }}
\rhead{{\bf \color{scipostdeepblue} ~Submission }}

\fancyfoot[C]{\textbf{\thepage}}
}

\usepackage{xspace}
\usepackage{booktabs}
\usepackage{multirow}
\usepackage[section]{placeins}   %

\newsavebox\DataMCRefBox
\newsavebox\DataMCBox
\newcommand{\datamcgraphics}[1]{%
  \sbox\DataMCRefBox{\includegraphics{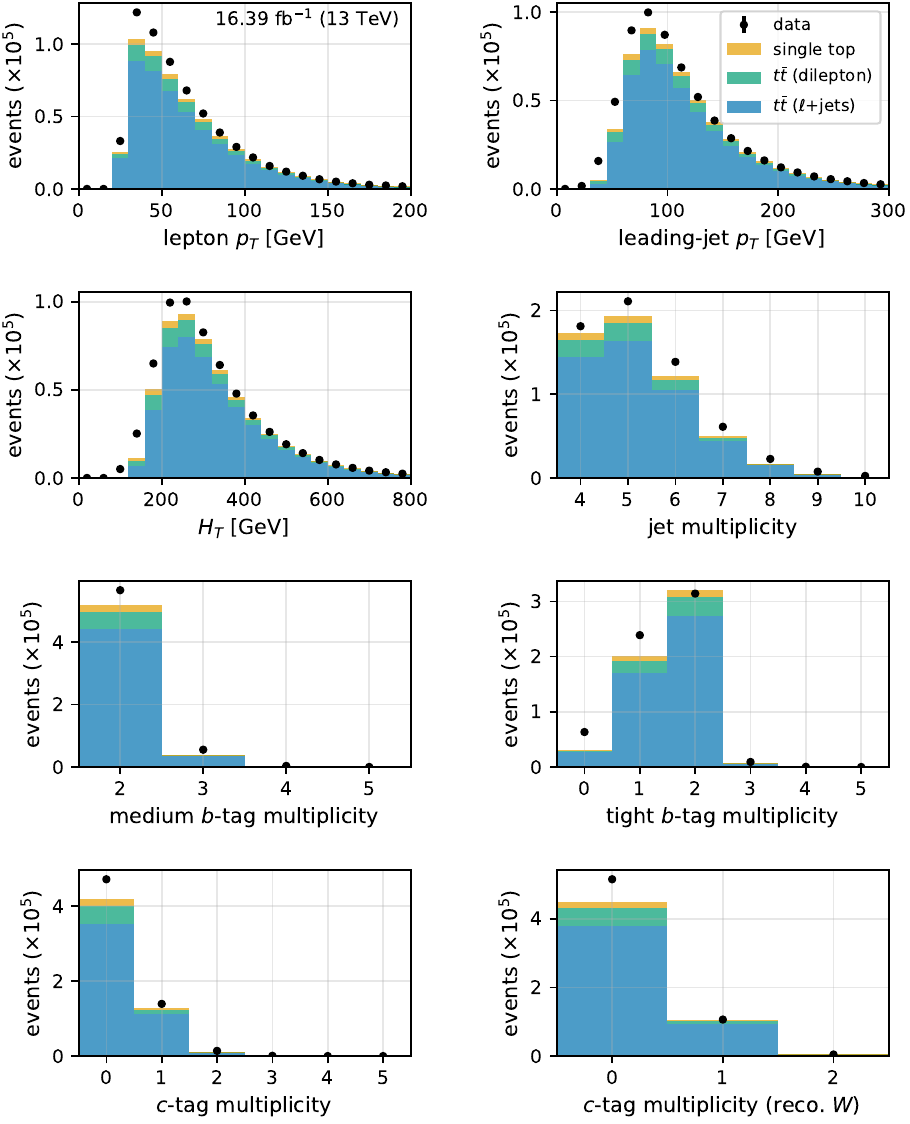}}%
  \sbox\DataMCBox{\includegraphics{#1}}%
  \includegraphics[width=\dimexpr0.828\textwidth*\wd\DataMCBox/\wd\DataMCRefBox\relax]{#1}}
\input{note_numbers.tex}

\newcommand{\pt}{\ensuremath{p_{\mathrm{T}}}\xspace}
\newcommand{\GeV}{\text{GeV}}
\newcommand{\MeV}{\text{MeV}}
\newcommand{\TeVns}{\text{TeV}}
\newcommand{\ttbar}{\ensuremath{t\bar{t}}\xspace}
\newcommand{\DR}{\ensuremath{\Delta R}\xspace}
\newcommand{\lamhat}{\hat{\Lambda}}%

\newcommand{\PaperTitle}{A first look at strange-quark polarization in top-quark decays with CMS Open Data}
\newcommand{\PaperAuthors}{%
\begin{center}\textbf{
Yevgeny Kats
}\end{center}

\begin{center}
Department of Physics, Ben-Gurion University, Beer-Sheva 8410501, Israel
\\[\baselineskip]
\href{mailto:katsye@bgu.ac.il}{\small katsye@bgu.ac.il}
\end{center}}
\newcommand{\eprint}[2][]{\href{https://arxiv.org/abs/#2}{arXiv:#2\if\relax\detokenize{#1}\relax\else\ [#1]\fi}}

\begin{document}

\pagestyle{SPstyle}

\begin{center}{\Large \textbf{\color{scipostdeepblue}{
\PaperTitle\\
}}}\end{center}

\PaperAuthors

\section*{\color{scipostdeepblue}{Abstract}}
\textbf{\boldmath{%
We present a first realization of strange-quark polarimetry in top-quark decays. The
longitudinal polarization of $\Lambda$ baryons is measured in the strange jets of hadronic $W\to c\bar s$
decays in semileptonic top-quark pair events, using $\LumiDet\,\text{fb}^{-1}$ of the public CMS data of
pp collisions at $\sqrt{s}=13\,\TeVns$. The $W$ decay produces the strange quark fully polarized, so that
the $\Lambda$ polarization measures the transfer of the quark polarization in fragmentation. With the
limited statistics available, our result is consistent both with zero and with the transfer inferred from
the flavor-averaged LEP measurements. The statistical reach with the full Run-2, Run-3 and HL-LHC datasets is
projected.
}}

\vspace{\baselineskip}

\enlargethispage{2\baselineskip}
\vspace{10pt}
\noindent\rule{\textwidth}{1pt}
\tableofcontents
\noindent\rule{\textwidth}{1pt}
\vspace{10pt}

\section{Introduction}
\label{sec:intro}

How the spin of a quark is transferred to the hadrons it fragments into is among the least
tested aspects of QCD. The spin-dependent fragmentation functions are nonperturbative and
weakly constrained. However, a polarized quark of known polarization turns its leading baryon into a
direct experimental probe of them, which motivates the work we present here. The question is also of
practical importance as it can allow measuring quark polarization in any given sufficiently large sample.

The momentum of an energetic quark is readily estimated from the jet it
produces, but its polarization is far harder to access. The spin information can be
degraded, or washed out entirely, during the nonperturbative fragmentation into hadrons;
the hadron that carries the original quark is generally not identifiable among the many
produced in the jet; and the polarization can be lost in the subsequent decay chain, before
any parity-violating decay that would reveal it. For heavy quarks the
situation is more favorable. The quark's chromomagnetic moment scales as $1/m_q$, so its spin
is protected during hadronization. The heavy-flavored hadron is identifiable since it usually carries a large fraction of the jet energy
and decays at a displaced vertex, and the spin of the $b$ or $c$ quark is protected inside the $\Lambda_b$ or $\Lambda_c$
baryon throughout its lifetime. The polarizations can then be measured through baryon
decays~\cite{falk-peskin}.
For the strange quark, whose mass is of order $\Lambda_{\rm QCD}$, no such protection is
guaranteed, and \emph{a priori} the polarization need not survive fragmentation at all.
Measurements at LEP nonetheless established that energetic $\Lambda$ baryons retain an
$\mathcal{O}(1)$ fraction of the strange-quark longitudinal polarization. In $Z\to s\bar s$ decays,
the strange quark is expected to be produced with a longitudinal polarization
$P(s)\approx-0.94$.\footnote{This is the Born-level value,
$P(s)=-2v_sa_s/(v_s^2+a_s^2)$ with $v_s=-\tfrac12+\tfrac23\sin^2\theta_{\mathrm{eff}}$ and $a_s=-\tfrac12$,
evaluated at the effective weak mixing angle $\sin^2\theta_{\mathrm{eff}}=0.2315$. The
$\mathcal{O}(\alpha_s)$ correction is an anomalous helicity-flip effect that survives the $m_q\to0$
limit~\cite{korner-oneloop}, multiplying the phase-space-inclusive polarization by
$(1+\alpha_s/3\pi)/(1+\alpha_s/\pi)\simeq1-2\alpha_s/3\pi\approx0.975$, to $\approx-0.91$.
However, the helicity-flipped quarks are soft, with spectrum $\propto(1-x)$ in the quark energy
fraction~\cite{falk-sehgal}, so for a leading baryon carrying a fraction $z>0.3$ of the
beam momentum the depolarization is suppressed by at least $(1-z)^2$, and the
polarization lies between $-0.92$ and the Born value.\label{fn:psz}} ALEPH and OPAL measured the
polarization of $\Lambda$ baryons in inclusive hadronic $Z$
decays~\cite{aleph-buskulic,aleph-lambdapol,opal-lambdapol} and found, for the $\Lambda$
momentum fraction $z\equiv p^{\,\Lambda}/p^{\,\text{beam}}>0.3$ (approximately the jet
momentum fraction for two-jet $Z$ events), $P(\Lambda)=-0.31\pm0.05$
(ALEPH~\cite{aleph-lambdapol}) and $-0.33\pm0.08$ (OPAL~\cite{opal-lambdapol}), with a
larger magnitude for the harder cut $z>0.4$.

The LEP measurements were \emph{inclusive},
averaging over all quark flavors --- comparable numbers of $u$, $d$, $s$, $c$, and $b$
jets --- so $\Lambda$ baryons from non-strange jets also enter the quoted
numbers. $\Lambda$ baryons carrying a primary $u$ or $d$ quark may retain a fraction of that
quark's polarization, of the \emph{opposite} sign (the $u$ and $d$ carry
$\Lambda$ spin oppositely to the $s$, as inferred from nucleon spin-structure data), while $\Lambda$ baryons from
$c$ and $b$ jets contribute with their own polarizations~\cite{burkardt-jaffe,jaffe-lambdafrag,kotzinian-bravar,boros-liang}.
Within the strange jets themselves, $\Lambda$ baryons from the decays of heavier strange baryons
($\Sigma^0$, $\Sigma^{\ast\,\pm,0}$, $\Xi^{0,-}$) --- a contribution comparable to direct $\Lambda$
production --- and from secondary strange quarks dilute the signal further.
Once this flavor composition is unfolded, the LEP
simulations indicate~\cite{kats-spol,gustafson-hakkinen,aleph-lambdapol,opal-lambdapol} that a pure sample of strange jets, with its strange-sector dilutions,
would have $P(\Lambda)\approx-\PLamBandLo$ to
$-\PLamBandHi$ for $z>0.3$, i.e.\ a large transfer of the strange-quark polarization to the
$\Lambda$.\footnote{The LEP analyses extracted $P(\Lambda)$ from the measured decay asymmetry using the
then-current $\alpha_\Lambda=\AlphaLEP$, and the unfolded pure-strange-jet band was correspondingly
quoted as $-0.49$ to $-0.65$. Since the data fix the product $\alpha_\Lambda P(\Lambda)$ and
not $P(\Lambda)$ itself, we restate the band in the present convention
$|\alpha_\Lambda|=\AlphaL$ (discussed below), which rescales it by $\AlphaLEP/\AlphaL$.}
Because the fragmentation factorizes from the hard process, this transfer
applies --- up to calculable scale evolution~\cite{dfsv} --- to strange quarks produced in any
process at a comparable energy scale.

The polarization is read out from the parity-violating weak decay $\Lambda\to p\pi^-$
(branching fraction $64\%$~\cite{pdg}), which produces a displaced two-prong vertex
($c\tau_\Lambda\approx7.9\,$cm) that reconstructs the $\Lambda$ mass. Its angular distribution,
\begin{equation}
\frac{d\Gamma}{d\cos\theta}\;\propto\;1+\alpha_\Lambda\,P(\Lambda)\cos\theta\,,
\label{eq:decay}
\end{equation}
relates the proton emission angle $\theta$ in the $\Lambda$ rest frame, measured relative to
any selected axis, to the polarization component along that axis through the decay asymmetry parameter
$\alpha_\Lambda=0.746\pm0.008$~\cite{pdg}. The conjugate decay $\overline\Lambda\to\bar p\pi^+$ has
$\alpha_{\overline\Lambda}=-0.758\pm0.005$~\cite{pdg}, opposite in sign and consistent in magnitude.
Assuming CP symmetry ($\alpha_{\overline\Lambda}=-\alpha_\Lambda$), we combine the two into a single
analyzing power $|\alpha_\Lambda|=\AlphaL$ --- their inverse-variance average --- used throughout.
The same weak decay was used by ATLAS to measure the \emph{transverse} polarization
of \emph{inclusively} produced $\Lambda$ at the LHC, found consistent with zero~\cite{atlas-lambdatrans}.
A transverse polarization referenced to the jet axis has later been observed by Belle in
$e^+e^-$ fragmentation~\cite{belle-lambdatrans}, enabling first extractions of the
corresponding T-odd ``polarizing'' fragmentation functions~\cite{dalesio-pff,callos-pff}. The
present measurement instead probes the \emph{longitudinal} polarization transferred from a \emph{polarized} strange quark.

As was proposed in ref.~\cite{kats-spol}, ATLAS and CMS can measure the polarization
of the longitudinally polarized strange quarks (with
$P(s)\simeq-1$)\footnote{The $\mathcal{O}(\alpha_s)$ helicity-flip correction of footnote~\ref{fn:psz}
multiplies the inclusive polarization to $\approx-0.98$; its suppression for leading baryons
brings the effective value back to $-0.99$ to $-1$.\label{fn:psw}} produced in $W^+\to c\bar s$
decays in single-lepton \ttbar events.
Beyond providing a benchmark that both validates the technique and constrains
the spin-dependent fragmentation, the \ttbar topology offers handles unavailable in the
flavor-inclusive LEP measurements: the charge of the lepton from the other top fixes whether a
$\Lambda$ or a $\overline\Lambda$ is expected on the hadronic side, halving the contribution of
secondary $\Lambda$ baryons; and charm tagging separates $W^+\to c\bar s$ from
$W^+\to u\bar d$ and identifies the strange jet as the $W$ jet recoiling against the tagged charm
jet. Moreover, already in LHC Run 2 the \ttbar yield of polarized strange quarks in ATLAS/CMS was
competitive with the $Z$ samples collected by the individual LEP experiments~\cite{kats-spol}.\footnote{As an aside, we note that \ttbar events enable
measurements in polarized quark samples of other flavors as well: the bottom quark through the $\Lambda_b$
baryon~\cite{galanti-heavybaryon}, the charm quark through the $\Lambda_c$~\cite{galanti-heavybaryon}, and
the up quark through the $\Sigma^+$ (which also provides a second handle on the strange)~\cite{kats-sigma}.}
The quantities at stake are the longitudinally polarized $\Lambda$ fragmentation functions
$\Delta D_{\Lambda/q}(z)$, for which the only global
information remains the LEP-era analysis of ref.~\cite{dfsv}.

In this paper we present a first realization of this measurement. We use a dataset with an
integrated luminosity of $\LumiDet\,\text{fb}^{-1}$ at $\sqrt{s}=13\,\TeVns$ from the CMS Open Data.
After a standard single-lepton \ttbar selection, we reconstruct the hadronic $W^+\to c\bar s$ decay, identify the strange jet by charm-tagging its
partner, reconstruct $\Lambda\to p\pi^-$ decays within it, and extract the polarization from the
decay angle in the high-$z$ ($z>0.3$) region where the polarization is expected to be sizable. The value of the
present work lies in establishing the full analysis chain end-to-end on collision data, quantifying the
dominant dilutions, and identifying the systematics that a precise measurement with the larger datasets
that are already available in CMS and ATLAS must control.

The paper is organized as follows. Section~\ref{sec:sample} describes the collision data and the
simulated samples. Section~\ref{sec:selection} presents the event selection, the reconstruction of the
hadronic top quark, and the identification of the charm and strange jets. Section~\ref{sec:v0}
describes the reconstruction and selection of the $\Lambda$ candidates in the strange jet.
Section~\ref{sec:strange} defines the decay angle and the polarimeter axes,
describes the acceptance correction and the estimator, and presents the results, the cross-checks and
the systematic uncertainty, and the projections for larger datasets.
Section~\ref{sec:conclusions} presents our conclusions.

\section{Samples}
\label{sec:sample}

\subsection{Collision data}
\label{sec:datasample}

The measurements use 2016 proton--proton collision data recorded by CMS at
$\sqrt{s}=13\,\TeVns$ and published on the CERN Open Data portal~\cite{opendata-data}.
Two single-lepton primary datasets from data-taking eras Run2016G and Run2016H are used,
\texttt{SingleMuon} and \texttt{SingleElectron}, reconstructed under the
ultra-legacy (UL) 2016 conditions. Events are selected online by a single-lepton
high-level trigger (HLT): in the muon channel the unprescaled isolated-muon trigger
\texttt{HLT\_IsoMu24} (logically OR-ed with its tracker-muon variant
\texttt{HLT\_IsoTkMu24}), with a $24\,\GeV$ online \pt threshold; in the electron
channel the single-electron trigger \texttt{HLT\_Ele27\_WPTight\_Gsf}, with a
$27\,\GeV$ threshold and a tight identification applied online.

The analysis runs on the detailed (MiniAOD) data format, which retains the per-particle
tracks and the reconstructed $V^0$ (strange-hadron) vertices. The two datasets, \texttt{SingleMuon} and \texttt{SingleElectron}, cover the same
certified lumisections, $\LumiDet\,\text{fb}^{-1}$ in total.\footnote{Obtained
from the Open Data luminosity record with \textsc{brilcalc}
and the physics normalization tag of the record, as the per-lumisection recorded
luminosity summed over the certified lumisections of the analyzed files.} The
more compact, high-level (NanoAOD) edition of the same data supplies the event-selection cutflow of
table~\ref{tab:cutflow}, thus providing a cross check on the event selection.

\subsection{Simulated samples}
\label{sec:mcsample}

The signal process is modeled with the official CMS simulated sample of
semileptonic \ttbar production, published on the same Open Data
portal~\cite{opendata-record}. The hard process, $pp\to t\bar{t}$, is generated at next-to-leading order (NLO) in
quantum chromodynamics (QCD) with \textsc{Powheg}~v2 (\textit{hvq})~\cite{powheg-hvq}, with one $W$~boson
required to decay leptonically ($W\to\ell\nu$, $\ell=e,\mu,\tau$) and the other
hadronically. Parton showering, hadronization, and the underlying event are
modeled with \textsc{Pythia}~8 using the CP5 tune~\cite{pythia8,cp5}. Pileup is
overlaid with a multiplicity matching the 2016 conditions, the detector response
is simulated with \textsc{Geant4}, and the events are reconstructed under the same
ultra-legacy 2016 conditions as the data. The main characteristics are summarized
in table~\ref{tab:sample}. The signal is normalized with the
NNLO+NNLL inclusive \ttbar cross section, $831.76\,$pb ($m_t=172.5\,\GeV$)~\cite{toppp},
times the semileptonic branching fraction $0.438$ (including $\tau$), giving
$\sigma_{t\bar t\to\ell+\text{jets}}\approx364\,$pb. These samples do not simulate the baryon
polarization. We obtain polarized samples by reweighting the simulated events
according to the corresponding decay angular distributions.

\begin{table}[tp]
\centering
\begin{tabular}{lrccc}
\toprule
Process & $\sigma$ [pb] & Records & Events [$10^6$] & $\mathcal{L}_{\mathrm{eff}}$ [fb$^{-1}$] \\
\midrule
\ttbar\ ($\ell$+jets)  & $364$  & 67992, 67993 & $\NMCmini$   & $\LMCfb$      \\
\ttbar\ ($2\ell$)      & $88$   & 67800, 67801 & $\DlNev$     & $\DlLeff$     \\
$tW$, top              & $35.9$ & 64880, 64881 & $\StTWtNev$  & $\StTWtLeff$  \\
$tW$, antitop          & $35.9$ & 64824, 64825 & $\StTWatNev$ & $\StTWatLeff$ \\
$t$-channel, top       & $136$  & 64758, 64759 & $\StTctNev$  & $\StTctLeff$  \\
$t$-channel, antitop   & $81$   & 64658, 64659 & $\StTcatNev$ & $\StTcatLeff$ \\
\bottomrule
\end{tabular}
\caption{Simulated samples: the semileptonic \ttbar\ signal and the dileptonic
\ttbar\ and single-top backgrounds, with the inclusive cross section, the Open Data
record numbers of the MiniAOD (analysis) and NanoAOD (validation) editions, the number of
events analyzed, and the effective integrated luminosity $\mathcal{L}_{\mathrm{eff}}=N/\sigma$. All samples are generated with
\textsc{Powheg}+\textsc{Pythia}~8. The \textsc{Powheg}
$t$-channel single-top samples are interfaced to \textsc{MadSpin} for the top decay.}
\label{tab:sample}
\end{table}

The dileptonic \ttbar process, in which both $W$ bosons decay leptonically
($t\bar t\to2\ell2\nu$, $\ell=e,\mu,\tau$), is included as a background. It is the same
\textsc{Powheg}~v2 (NLO) ${}+{}$ \textsc{Pythia}~8 (CP5 tune) \ttbar production as the
signal, reconstructed under the same UL2016 conditions~\cite{opendata-dilepton}
(table~\ref{tab:sample}). It enters the one-isolated-lepton selection when one of the
two leptons is lost --- outside the acceptance or failing the isolation requirement ---
or when $W\to\tau\nu$ and the $\tau$ decays hadronically, so the event mimics the lepton+jets
topology.

A single-top simulation is used as a further background. It combines the two
dominant single-top modes --- $tW$ associated production and the $t$~channel, top
and antitop --- each generated at NLO with \textsc{Powheg}~v2, showered with
\textsc{Pythia}~8 (CP5 tune)~\cite{opendata-singletop} (table~\ref{tab:sample}); the
$t$-channel samples are interfaced to \textsc{MadSpin} to model the top-quark decay. The
much smaller $s$~channel ($\sigma\approx10\,$pb) is neglected.

The per-object efficiency scale factors and the pileup reweighting are applied to the
simulation as multiplicative per-event weights, built from the publicly released CMS
Physics Object Group correction files for the ultra-legacy 2016 (post-VFP)
conditions.
The pileup weight matches the simulated pileup profile to the one measured in data; the
muon and electron identification and isolation efficiencies~\cite{muon-cms,ele-cms} (and, for muons,
the trigger efficiency)\footnote{The electron trigger (\texttt{Ele27\_WPTight}) scale factor is not
applied because it is not included in the public release.} and the
$b$-tagging efficiency~\cite{btag-cms} are each corrected to their data-measured values,
the latter through the fixed-working-point prescription, with binned per-flavor tagging-efficiency maps taken from the
signal simulation. The $c$-tagging efficiency is corrected with a shape calibration; it is described with the $c$-tag requirement in
section~\ref{sec:hadtop}.

\section{Event selection and reconstruction}
\label{sec:selection}

\subsection{Physics objects}
\label{sec:objects}

Muons and electrons are required to lie on the single-lepton-trigger plateau,
\begin{equation}
\begin{aligned}
\mu\colon&\quad \pt > 26\,\GeV,\quad |\eta| < 2.4 ,\\
e\colon&\quad \pt > 30\,\GeV,\quad |\eta_{\mathrm{SC}}| < 2.5 ,
\end{aligned}
\end{equation}
and to pass the official CMS identification --- tight muons and tight cut-based electrons, the
latter outside the ECAL barrel--endcap transition region in supercluster pseudorapidity,
$1.444<|\eta_{\mathrm{SC}}|<1.566$.

Only isolated leptons are considered. For muons the relative isolation --- the scalar \pt\ sum of the other particles within
$\DR<0.4$ of the muon, divided by the muon \pt, in the standard pileup-corrected
(``$\Delta\beta$'') form~\cite{muon-cms} --- is required to satisfy $I_{\text{rel}}<0.15$. The cut-based
electron identification imposes its own isolation internally~\cite{ele-cms}.

Jets are clustered from the reconstructed particles with the
anti-$k_{\mathrm{T}}$ algorithm~\cite{antikt} with distance parameter $R=0.4$.
Charged-hadron subtraction removes charged pileup before clustering, and the
standard jet energy corrections~\cite{jec-cms} are then applied; the corrected jets are required to have
\begin{equation}
\pt > 20\,\GeV, \qquad |\eta| < 2.4 .
\end{equation}
Since every isolated lepton is also reconstructed as a jet, jets overlapping with
an isolated lepton within $\DR<0.4$ are removed. The
official tight jet identification~\cite{pf-cms}, which rejects detector noise and
spurious non-collision jets, is also applied.

Jets are tagged as $b$ or $c$ jets with the \textsc{DeepJet} multivariate
discriminant~\cite{deepjet,btag-cms}. The $b$ tag uses the \textsc{DeepJet} $b$
discriminant at its medium working point (discriminant $>0.2489$ for UL2016, a
nominal $b$-jet efficiency of about $80\%$). The $c$-tagging
working point is described in section~\ref{sec:hadtop}.

\subsection{Basic event selection}
\label{sec:basicsel}

The signal region targets the semileptonic \ttbar topology by requiring:
\begin{enumerate}
\item exactly one isolated lepton (muon or electron) on the single-lepton-trigger
plateau (section~\ref{sec:objects}), with the corresponding single-lepton trigger having fired;
\item at least four jets with $\pt>20\,\GeV$ and $|\eta|<2.4$;
\item at least two of these jets medium $b$-tagged.
\end{enumerate}

\subsection{Hadronic top-quark reconstruction}
\label{sec:hadtopreco}

\begin{figure}[tp]
\centering
\includegraphics[width=\textwidth]{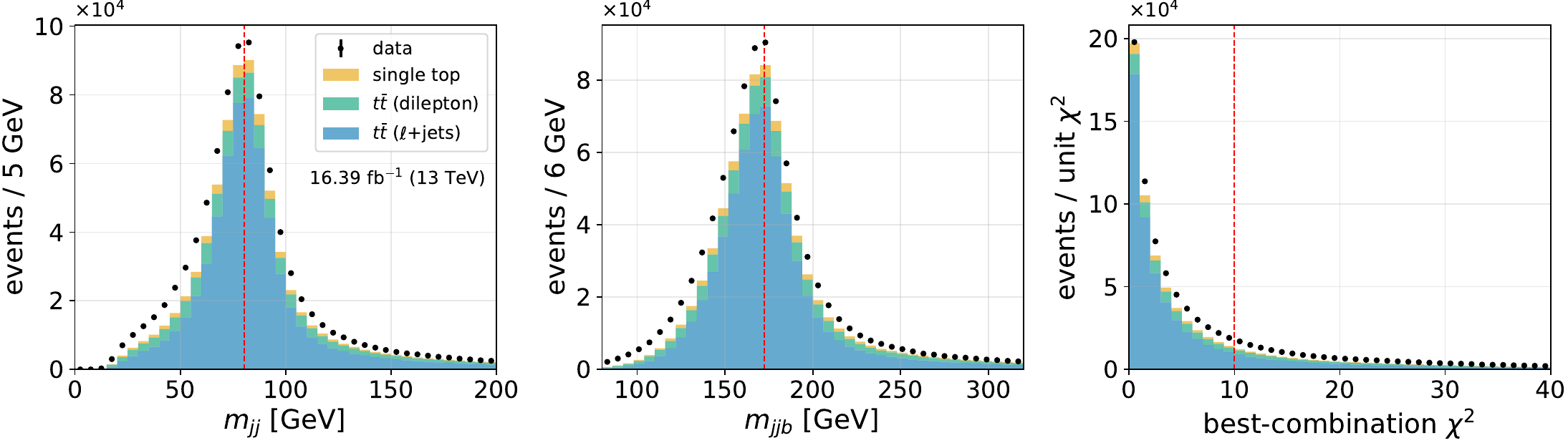}
\caption{Hadronic top reconstruction by $\chi^2$ minimization, eq.~\eqref{eq:chi2}, in
collision data (points) and in the stacked simulation (filled: the semileptonic
\ttbar signal, dileptonic \ttbar, and single top), with the simulation normalized to the
data luminosity. Panels, from left: the reconstructed hadronic $W$ mass $m_{jj}$ and top
mass $m_{jjb}$ (input masses dashed), and the best-combination $\chi^2$ with the
$\chi^2<10$ selection marked, all in the one-lepton, $\ge$4-jet, $\ge$2-$b$-tag
signal region.}
\label{fig:hadtop}
\end{figure}

In addition to the basic requirements above, the signal region requires a reconstructed
hadronically decaying top quark. The jet triplet combinatorics use the six highest-$\pt$ selected jets in the event. Among all jet triplets --- every $b$-tagged jet paired with every pair of non-$b$-tagged jets --- we take the one that minimizes
\begin{equation}
\chi^2 = \left(\frac{m_{jj}-m_W}{\sigma_W}\right)^2
       + \left(\frac{m_{jjb}-m_t}{\sigma_t}\right)^2 ,
\label{eq:chi2}
\end{equation}
with $m_W=80.4\,\GeV$ and $m_t=172.5\,\GeV$. The resolutions
$\sigma_W=\SigW\,\GeV$ and $\sigma_t=\SigTop\,\GeV$ are obtained from the simulation, as
the approximate Gaussian widths of the reconstructed $m_{jj}$ and $m_{jjb}$ peaks
for correctly (generator-)matched triplets.
Events whose best triplet has $\chi^2>10$ are vetoed.

The best triplet satisfies the
$\chi^2<10$ requirement in $\ChiSqSim\%$ of the simulated semileptonic \ttbar events that have at least one $jjb$ triplet, where the reconstructed
$W$ and top mass spectra (figure~\ref{fig:hadtop}) peak at the expected values, with
median masses of $\MjjMedSim$ and $\MjjbMedSim\,\GeV$ on a broad combinatorial background. Across the full simulation stack --- semileptonic and
dileptonic \ttbar and single top --- $\ChiSqSimFull\%$ of events with a triplet satisfy the
requirement. In the
collision data, $\ChiSqData\%$ of events with a triplet satisfy $\chi^2<10$ and the median reconstructed top mass is $\MjjbMedData\,\GeV$
(figure~\ref{fig:hadtop}, points).

\begin{table}[tp]
\centering
\setlength{\tabcolsep}{5pt}   %
\begin{tabular}{lrrrrrr}
\toprule
 & \multicolumn{4}{c}{Simulation (norm.\ to data luminosity)} & \multicolumn{1}{c}{\multirow{2}{*}{Data}} & \multicolumn{1}{c}{\multirow{2}{*}{\Large$\frac{\mathrm{Data}}{\mathrm{Sim.}}$}} \\
\cmidrule(lr){2-5}
Requirement & \multicolumn{1}{c}{$t\bar t\,(\ell{+}\mathrm{j})$} & \multicolumn{1}{c}{$t\bar t\,(2\ell)$} & \multicolumn{1}{c}{Single $t$} & \multicolumn{1}{c}{Total} & & \\
\midrule
\Cutflowbodynano
\bottomrule
\end{tabular}
\caption{Event-selection cutflow in data and simulation,
from the single-lepton requirement to the hadronic-top ($\chi^2<10$) signal-region condition.
Each row includes all preceding requirements. The simulated yields --- \ttbar\
($\ell$+jets) signal, dileptonic \ttbar, and single top --- are normalized to the data
luminosity ($\LumiDet\,\text{fb}^{-1}$). The $W$+jets and
QCD multijet backgrounds are not simulated and dominate the loose upper rows. The last column is
the data-to-simulation ratio.}
\label{tab:cutflow}
\end{table}

The cutflow through this selection is given in table~\ref{tab:cutflow}.
The unsimulated $W$+jets and QCD multijet backgrounds dominate the data in the loose upper
rows, but are progressively suppressed by the jet and $b$-tag
requirements.

\subsection{Charm- and strange-jet identification}
\label{sec:hadtop}

Charm tagging is applied to jets from hadronic $W$ decays. It uses the two-dimensional \textsc{DeepJet} $c$-vs-light and $c$-vs-bottom
discriminants (\textsc{CvsL}, \textsc{CvsB}) formed from the per-jet \textsc{DeepJet}
flavor probabilities. A jet is $c$-tagged when both discriminants exceed chosen
thresholds. A custom working point is used, with the thresholds defined in
table~\ref{tab:cwp}. The working point is validated against the generator-level jet flavor
that is based on ghost association of heavy-flavored hadrons~\cite{btag-cms}. Measured in
simulation without scale factors, over the signal-region jets before $b$ tagging (events with one
isolated lepton and at least four jets), it gives a $c$-jet efficiency of $\CeffMed$ with light- and $b$-jet mistag rates of
$\LmisCm$ and $\BmisCm$.

\begin{table}[htbp]
\centering
\begin{tabular}{lcc}
\toprule
Working point & \textsc{CvsL} & \textsc{CvsB} \\
\midrule
This analysis & $0.2$  & $0.4$  \\
\midrule
Official \textsc{DeepJet}, loose  & $0.039$ & $0.305$ \\
Official \textsc{DeepJet}, medium & $0.099$ & $0.353$ \\
Official \textsc{DeepJet}, tight  & $0.269$ & $0.247$ \\
\bottomrule
\end{tabular}
\caption{Charm-tagging working point on the two-dimensional \textsc{DeepJet}
$(\textsc{CvsL},\textsc{CvsB})$ discriminant. The first row describes the custom thresholds
defined for this analysis. The last three rows are the
official CMS \textsc{DeepJet} $c$-tag working points, shown for comparison. A jet passes a point
when both discriminants exceed the listed values.}
\label{tab:cwp}
\end{table}

\begin{figure}[tp]
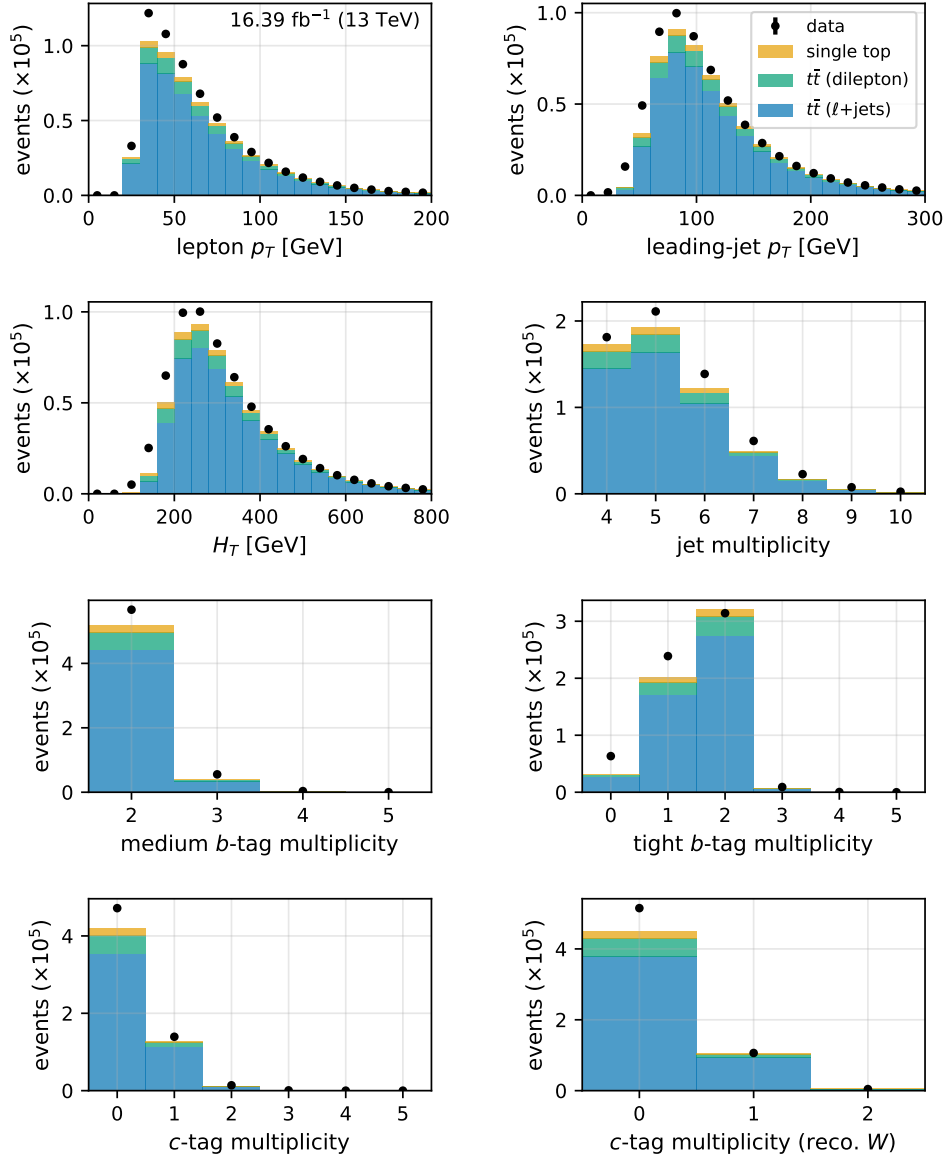

\centering
\datamcgraphics{plots/datamc_kinematics_multiplicities_mini.pdf}
\caption{Data/simulation comparison in the one-lepton, $\ge$4-jet, $\ge$2-$b$-tag,
reconstructed-hadronic-top signal region in collision data (points, statistical
errors) and the stacked semileptonic and dileptonic \ttbar\ and single-top simulation
(normalized to the data luminosity). First row: the lepton \pt and the leading-jet \pt.
Second row: the scalar jet-\pt sum $H_{\mathrm T}$ and the jet multiplicity. Third row: the
medium- and the tight-working-point $b$-tag multiplicities. Fourth row: the $c$-tag
multiplicity over the signal-region jets that fail the medium $b$ tag, and restricted to
the two reconstructed-$W$ jets. The charm-tagging efficiency corrections of
section~\ref{sec:hadtop} are applied to the reconstructed-$W$ panel only.}
\label{fig:datamc}
\end{figure}

The charm-tagging efficiency in simulation is corrected with the iterative-fit shape
calibration of the \textsc{DeepJet} charm discriminants~\cite{ctag-cms}, which assigns each jet
a scale factor according to its hadron flavor and its \textsc{CvsL} and \textsc{CvsB} values.
The scale factors are applied to the two jets assigned to the hadronic $W$ boson.\footnote{These
jets fail the medium $b$ tag by construction, and the $b$-tagging scale factors of
section~\ref{sec:mcsample} already correct their yield. The charm-tagging shape calibration must
therefore change only the discriminant distributions. Its scale factors are normalized to
unit mean for each hadron flavor over the $W$ jets of the signal region, before the charm-tag
requirement, and a factor for each simulated sample restores its yield there. The normalization assumes
that the flavor composition of the region in data is that of the simulation after the $b$-tagging
correction. The same scale factors are applied to jets with a $\Lambda$ candidate, although the
displaced tracks of its decay can enter the tagger inputs.}

The hadronic-$W$ pair of the top triplet is required to contain exactly one jet passing this
working point: that jet is the charm-jet candidate, and its $W$ partner is the
strange-jet candidate.

\subsection{Data and simulation in the signal region}
\label{app:datamc}

In the one-lepton, $\ge$4-jet, $\ge$2-$b$-tag signal region with the hadronic-top
requirement, data and simulation agree across the principal kinematic distributions
and tagged-jet multiplicities (figure~\ref{fig:datamc}): the stacked
semileptonic and dileptonic \ttbar\ and single-top prediction
describes the bulk of each distribution, with shape deviations in the tails. As already shown in
table~\ref{tab:cutflow}, the total data yield exceeds the prediction by $\DataMCdev\%$,
which can be attributed to the residual unsimulated $W$+jets and QCD multijet backgrounds and
the cross-section, $b$-tagging, and luminosity uncertainties.

\section{\texorpdfstring{$\Lambda$}{Lambda}-baryon reconstruction and selection}
\label{sec:v0}

We search the strange jet --- the hadronic-$W$ jet recoiling against the $c$-tagged
jet --- for $\Lambda\to p\pi^-$ decays (and the charge conjugate
$\bar\Lambda\to\bar p\pi^+$), reconstructed as displaced $V^0$ vertex
candidates~\cite{cms-tracking-2010,cms-strangeprod} within $\DR<0.4$ of the jet axis.\footnote{We use the
\texttt{slimmedLambdaVertices} collection stored in the MiniAOD samples, which is produced
centrally by the standard CMS $V^0$ fitter under the $p\pi$ mass hypothesis.}
To label each candidate as a $\Lambda$ or $\bar\Lambda$ and to reconstruct the decay
angle, the daughters of the $V^0$ are used.\footnote{Their momenta are provided at the
track's point of closest approach to the primary vertex, and are corrected here for the
rotation the tracks undergo in the $3.8\,$T magnetic field between that point and the
$V^0$ decay vertex.} The proton is assumed to be the higher-momentum
daughter --- in simulation it is the correct choice in more than $\LamProtonFrac\%$
of the candidates that pass the full $\Lambda$ selection (inclusive in
$z$) --- and its charge labels the candidate a $\Lambda$ or a
$\bar\Lambda$. Candidates in which either daughter is identified as an electron or a muon by the particle-flow
reconstruction are rejected.

\begin{table}[tp]
\centering
\resizebox{\textwidth}{!}{%
\begin{tabular}{lrrrrrrc}
\toprule
 & \multicolumn{5}{c}{Simulation (normalized to data luminosity)} & \multicolumn{1}{c}{\multirow{2}{*}{Data}} & \multirow{2}{*}{Data/Sim.} \\
\cmidrule(lr){2-6}
Requirement & \multicolumn{1}{c}{$t\bar t\,(\ell{+}\mathrm{j})$} & \multicolumn{1}{c}{$t\bar t\,(2\ell)$} & \multicolumn{1}{c}{Single $t$} & \multicolumn{1}{c}{Total} & \multicolumn{1}{c}{\shortstack{True $\Lambda$ in\\true $s$ jet}} & & \\
\midrule
\Cutflowbodymini
\bottomrule
\end{tabular}%
}
\caption{Event-selection cutflow of the $\Lambda$ polarimetry signal region, continuing
table~\ref{tab:cutflow} from the hadronic-top ($\chi^2<10$)
step. The subsequent rows add the
single-$c$-tag requirement that defines the strange jet, and then the $\Lambda$
selection of this section one cut at a time: at least one
$\Lambda$ or $\bar\Lambda$ candidate in that jet; the $p\pi$ mass window; the veto of candidates
consistent with the $K_S\to\pi^+\pi^-$ hypothesis; and finally the retention of the highest-$p_T$
candidate whose baryon number is consistent with the expectation from the lepton charge. The two momentum-fraction ($z$) threshold rows complete the table.
The ``True $\Lambda$'' column is a subset of the $t\bar t\,(\ell{+}\mathrm{j})$ column. It counts the
candidates that are a genuine $\Lambda$ or $\bar\Lambda$, that is, a $V^0$ matched to a generator-level
$\Lambda$ or $\bar\Lambda$ within $\Delta R<0.1$, and that carry the baryon number expected from the
lepton charge. The candidate is required, in addition, to lie inside a genuine strange jet of the flavor
the lepton charge predicts: a jet of parton flavor $s$ for $\ell^+$ and $\bar s$ for $\ell^-$, whose
partner jet in the hadronic $W\to c\bar s$ decay is truth-charm.
The uncertainty quoted on the data-to-simulation ratio, in the rows where it is not
negligible, combines the statistical uncertainties of the data and simulated yields.}
\label{tab:minicutflow}
\end{table}

The $p\pi$ mass is required to be within $6\,$MeV of $m_\Lambda$, which is about
${\pm}\LamMassNsig\sigma$ of the $\LamMassRes\,$MeV core mass resolution and retains
$\LamMassKeep\%$ of the genuine (generator-matched) $\Lambda$ candidates in simulation, the remainder falling in a broader tail from
poorly measured daughter tracks. In addition, a candidate is vetoed if, taking its two daughters as
pions, their invariant mass lies within $5\,$MeV of the observed $\pi^+\pi^-$ invariant mass peak
produced by true $K_S$ that are present in the $\Lambda$ collection in the simulated sample.\footnote{Computed from the daughter
momenta, this mass peaks $\KsVetoOffset\,$MeV below $m_{K_S}$, at $\KsVetoCtr\,$MeV.}

\begin{figure}[tp]
\centering
\includegraphics[width=\textwidth]{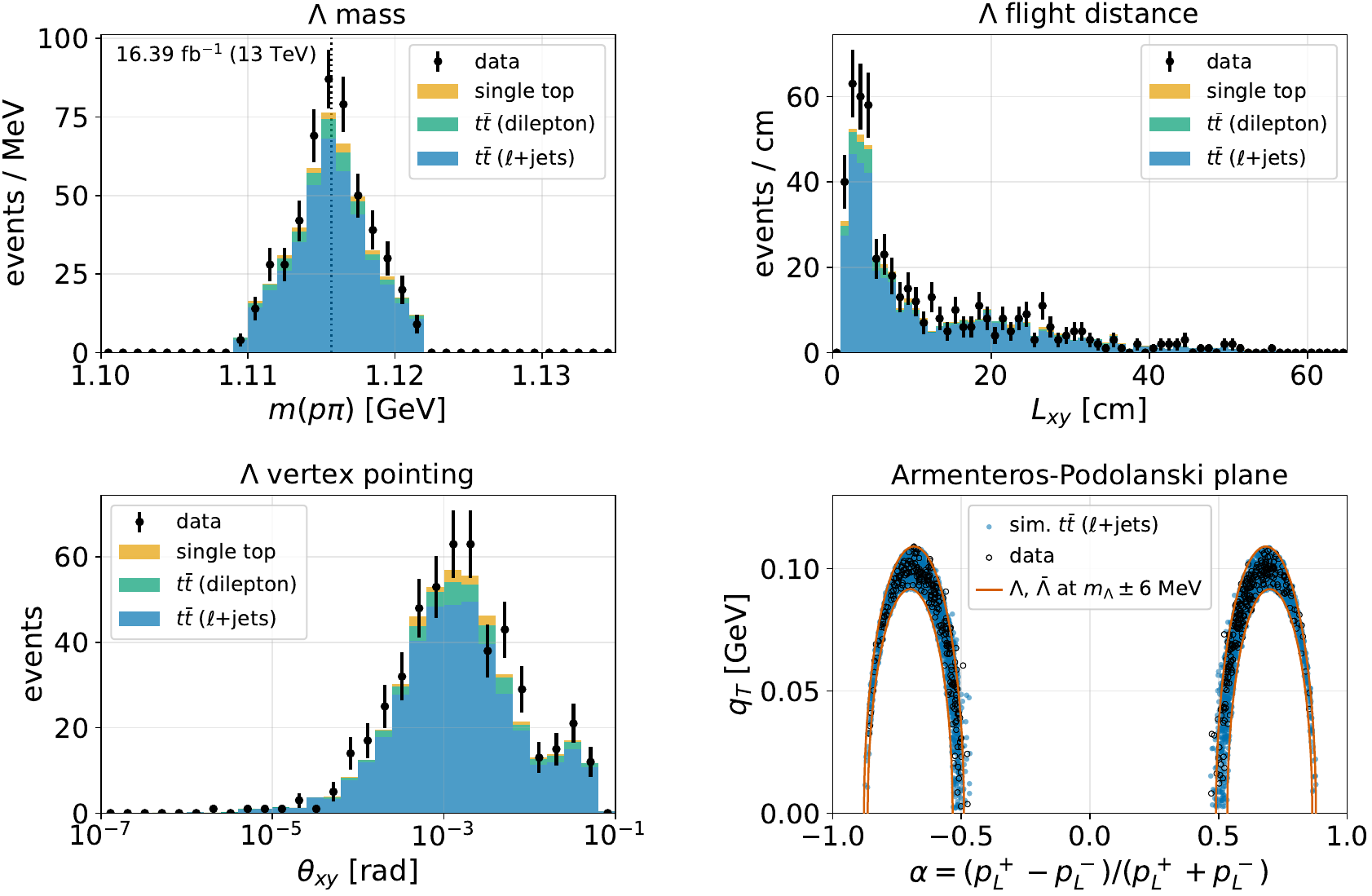}
\caption{Reconstruction cross-checks for the strange-jet $\Lambda$ sample, on the
selected candidates that enter the polarimetry, in collision data and
simulation. Top row: the $p\pi$ mass peak (PDG value dotted) and the transverse flight
distance $L_{xy}$; bottom left: the 2D vertex pointing angle $\theta_{xy}$. In these three
panels the simulation is the stack of all samples ---
semileptonic \ttbar, dileptonic \ttbar, and single top --- normalized to the data
luminosity. Bottom right: the Armenteros--Podolanski plane, where the candidates lie on the
asymmetric $\Lambda$ ($\alpha>0$) and $\bar\Lambda$ ($\alpha<0$) arcs. The solid curves are the
two-body kinematic ellipses of the $\Lambda\to p\pi^-$ and $\bar\Lambda\to\bar p\pi^+$ decays for a
relativistic candidate ($\beta\to1$) at the two edges of the $p\pi$ mass window, $m_\Lambda\pm6\,$MeV.
This panel is a
per-candidate scatter, shown for the data and the semileptonic \ttbar simulation.}
\label{fig:sjetquality}
\end{figure}

The surviving candidates are then
required to carry the baryon number expected from the lepton: because the charged lepton and the
strange jet originate from opposite top quarks, an $\ell^+$ (from $W^+\!\to\ell^+\nu$) implies a
hadronic $W^-\!\to\bar c\,s$, whose strange jet carries an $s$ quark and is therefore
expected to fragment to a $\Lambda$, whereas an $\ell^-$ implies $W^+\!\to c\bar s$ and a
$\bar\Lambda$. If multiple $\Lambda$ or $\bar\Lambda$ candidates in the jet satisfy the
above requirements, the highest-$\pt$ candidate is selected. This choice is seldom consequential:
of the jets holding at least one such candidate, only about 1\% hold more than one.

\begin{figure}[tp]
\centering
\includegraphics[width=\textwidth]{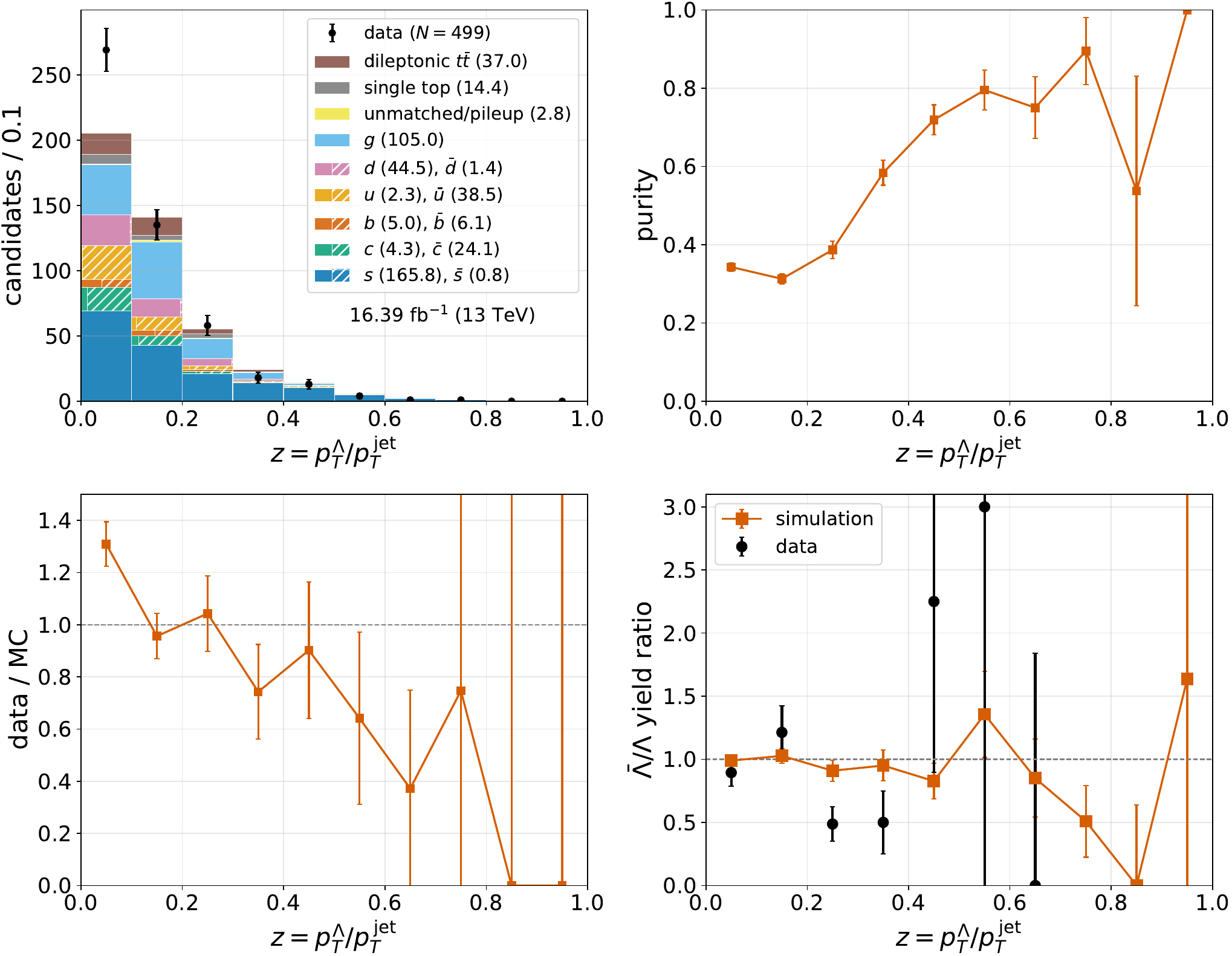}
\caption{The $\Lambda$ and $\bar\Lambda$ candidates in the strange-jet candidates as a function of the momentum
fraction $z=\pt^{\,\Lambda}/\pt^{\,\text{jet}}$. \emph{Top left:} collision data (points) and simulation
normalized to the data luminosity ($\LumiDet\,$fb$^{-1}$). The semileptonic \ttbar\ is stacked
by the generator flavor of the tagged jet, with single top (gray) and dileptonic \ttbar\ (brown)
as separate components. Each flavor component is split into two cases: the solid
part is the fraction whose $\Lambda$ baryon number sign agrees with that of the jet's parton
(quark jet $\to\Lambda$, antiquark jet $\to\bar\Lambda$), the hatched part the opposite pairing.
The legend labels the two yields by the parton that makes them in the $\Lambda$ convention, the
quark for the agreeing pairing and the antiquark for the opposite.
\emph{Top right:} the purity in the simulated sample: the fraction of the candidates
whose jet is a genuine strange jet with the agreeing pairing, versus $z$. \emph{Bottom left:} the data-to-simulation yield ratio versus $z$. \emph{Bottom right:} the $\bar\Lambda/\Lambda$
yield ratio versus $z$ in data (points) and simulation (squares).}
\label{fig:zsplit}
\end{figure}

The cutflow of the selection is given in table~\ref{tab:minicutflow},
continuing table~\ref{tab:cutflow}
from the hadronic-top ($\chi^2<10$) step.\footnote{The two tables are evaluated on different editions
of the same events: table~\ref{tab:cutflow} on NanoAOD, and table~\ref{tab:minicutflow} on the MiniAOD
samples that the analysis uses. The two
formats share the
hadronic-top row, where they agree to ${\sim}\CFShareDevData\%$ in data and to
${\sim}\CFShareDevSim\%$ in simulation. The residual is a threshold migration: NanoAOD stores the
jet momenta and the $b$-tagging discriminants with a truncated mantissa, so a jet within a
per-mille of a selection boundary can fall on either side of it in the two formats.}
Figure~\ref{fig:sjetquality} shows the basic properties of the selected sample of $\Lambda$
candidates: the $p\pi$ mass peak, the transverse flight distance $L_{xy}$, the
two-dimensional vertex pointing angle $\theta_{xy}$ between the $V^0$ momentum and its
$L_{xy}$ vector from the beam line, and the Armenteros--Podolanski plane: the daughter
momentum $q_T$ transverse to the candidate's flight direction versus the daughter
longitudinal-momentum asymmetry $\alpha=(p_L^+-p_L^-)/(p_L^++p_L^-)$. The
candidates populate the asymmetric $\Lambda$ ($\alpha>0$) and $\bar\Lambda$ ($\alpha<0$)
Armenteros arcs. The candidates are genuinely displaced, with a median
transverse flight distance $L_{xy}\approx\LxyMed$\,cm.

Figure~\ref{fig:zsplit} analyzes the $\Lambda$ candidates as a function of the momentum fraction
$z=\pt^{\,\Lambda}/\pt^{\,\text{jet}}$
carried by the $\Lambda$ within its jet.\footnote{This approximates the full-momentum fraction
$p^{\,\Lambda}/p^{\,\text{jet}}$ to the extent that the
$\Lambda$ is collinear with the jet axis. For the candidates passing $z>0.3$,
the full-momentum fraction averages
$\langle p^{\,\Lambda}/p^{\,\text{jet}}\rangle\approx\ZfullMeanZthree$, with
$\ZfullFracAbove\%$ of them above $0.3$ in the momentum variable as well.}
The top-left plot shows the $z$ distribution of $\Lambda$ and $\bar\Lambda$ in
strange-jet candidates, in collision data and in simulation normalized to the data luminosity and
split by the flavor of the tagged jet:\footnote{The parton flavor (including the
quark/antiquark distinction) of a jet is based here on the generator-level parton that the CMS jet-flavor
clustering associates with it. The generator partons are clustered into the jet as ghosts of
vanishing momentum~\cite{cacciari-salam-areas}.} the genuine
strange jet of $W^-\to s\bar c$ versus the charm, bottom, up, down, and gluon mis-tags. Within each quark flavor the candidates are
further divided by whether the jet containing the $\Lambda$ was initiated by a quark or an
antiquark. The $\bar\Lambda$ contributions are counted oppositely in this division.
Only the agreeing pairing --- a quark jet fragmenting to a $\Lambda$, an antiquark to a
$\bar\Lambda$ --- can transmit the parton polarization, while in the opposite one the $\Lambda$
draws an (anti)quark of the required flavor from the sea and keeps no memory of the parton spin.
Counting only the agreeing strange pairing, the genuine strange fraction is \PurMed\%.
The strange purity climbs
with $z$ (top right), high-$z$ $\Lambda$ baryons being the more likely to arise from the
fragmentation of a hard strange quark. The simulation roughly reproduces both the observed yield (bottom left) and the
$\bar\Lambda/\Lambda$ yield ratio (bottom right).

\begingroup\let\FloatBarrier\relax
\section{\texorpdfstring{$\Lambda$}{Lambda}-baryon polarimetry}%
\label{sec:strange}%
\label{sec:v0data}%
\endgroup

\subsection{Decay angle and polarimeter axes}
\label{sec:v0pol}

The $\Lambda$ and the proton candidate produced in its decay are first boosted into the
hadronic-$W$ rest frame, in which the $\Lambda$ flight direction defines the longitudinal axis
$\lamhat$. The proton is then boosted from that frame into the $\Lambda$ rest frame, and the decay
angle $\theta$ is taken between its momentum and $\lamhat$. Two transverse axes complete a
right-handed triad. Both are formed in the $W$ rest frame. With $\hat w$ the direction of the $W$ boson in the
top-quark rest frame, which in the $W$ rest frame is the direction opposite to the top-quark momentum,
the normal axis is defined by $\hat w\times\lamhat$, perpendicular to the plane spanned by $\hat w$ and
$\lamhat$, and the in-plane axis is along $\lamhat\times(\hat w\times\lamhat)$.

Equation~\eqref{eq:decay} gives $\langle\cos\theta\rangle=\alpha P/3$, with the analyzing power
$|\alpha|=\AlphaL$ of section~\ref{sec:intro}, so that $P=\pm1$, for example, correspond to
$\langle\cos\theta\rangle\approx\pm\CosFullP$. Throughout this section $P$ denotes the polarization of
the $\Lambda$ and \emph{minus} that of the $\bar\Lambda$, so that the proton and the antiproton
analyzers add. For the longitudinal polarization (helicity), a left-handed $\Lambda$ and a
right-handed $\bar\Lambda$, which is the expected consequence of the opposite $s$ and $\bar s$
helicities of the $W\to c\bar s$ decay, both correspond to negative $P$.

Both transverse components vanish for the signal. The $V-A$ coupling produces the strange quark of
$W\to c\bar s$ in a definite helicity state, so that a transverse quark polarization would require a
helicity flip, suppressed by $m_s/m_W$, and fragmentation, being independent of the production plane,
can polarize the $\Lambda$ only along the quark axis. A normal component would, in addition, be a T-odd
correlation requiring absorptive or CP-violating phases. The T-odd transverse polarization of
fragmentation hyperons~\cite{belle-lambdatrans} arises without any quark polarization and points along
the normal to the plane of the quark and $\Lambda$ momenta. Like any polarization tied to the $\Lambda$
momentum transverse to the quark, it rotates with the $\Lambda$ from candidate to candidate, because that
momentum is distributed uniformly in azimuth around the quark direction. Its projection on either
transverse axis used here then averages to zero over the sample, up to a residual of order
$k_T/p_\Lambda$, the ratio of the transverse to the full $\Lambda$ momentum: the axes are defined by the
$\Lambda$ direction, which differs from the quark direction by that angle. The transverse projections
therefore serve as null consistency checks and as in-situ monitors of the acceptance and background
distortion.

\subsection{Acceptance, reweighting, and templates}
\label{sec:v0acc}

\begin{figure}[tp]
\centering
\includegraphics[width=\textwidth]{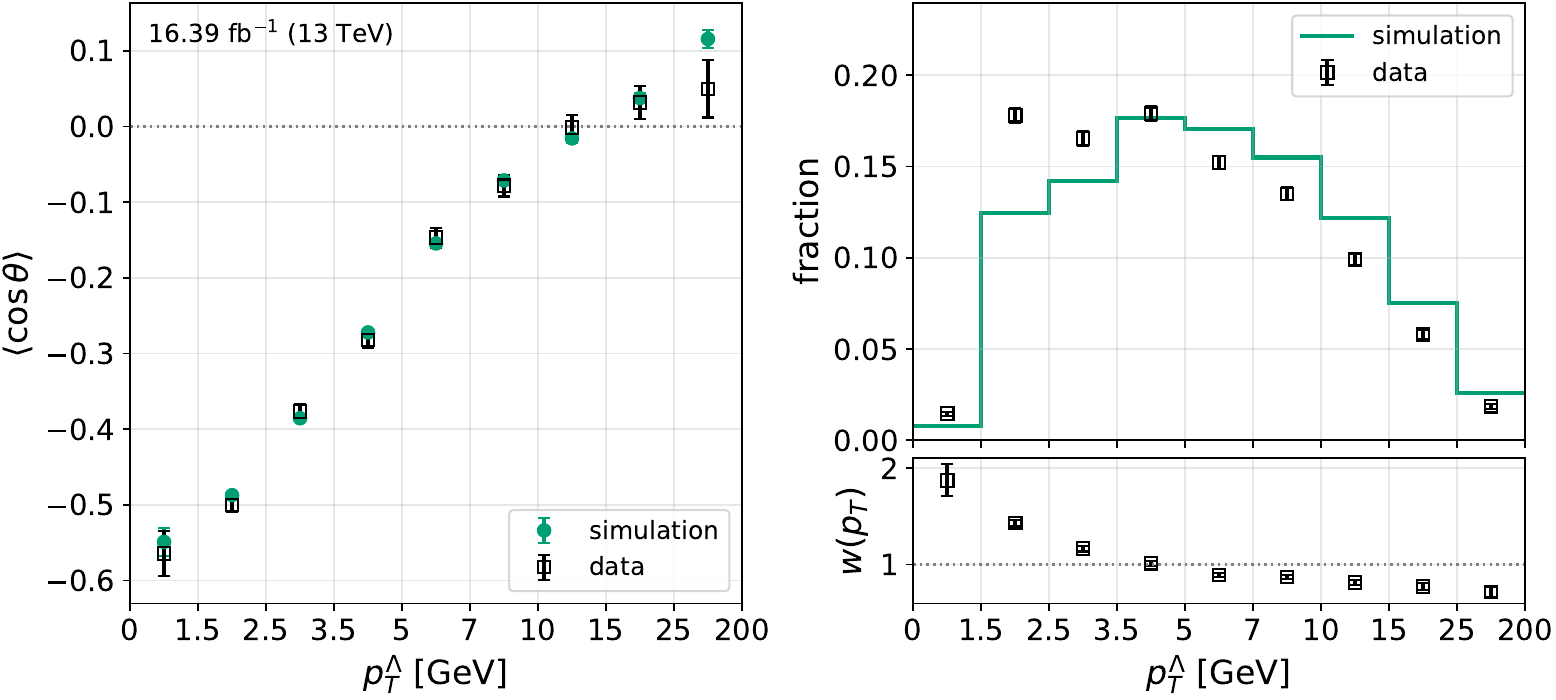}
\caption{The $\Lambda$-baryon $p_T$ dependence of $\langle\cos\theta\rangle$ and the $\Lambda$-baryon
$p_T$ spectrum in the control region of the $p_T$ reweighting, which comprises the $\Lambda$
candidates in any of the three jets of the top triplet. Left:
$\langle\cos\theta\rangle$ for the longitudinal axis versus $p_T^{\Lambda}$ for the unpolarized
simulation (filled) and the data (open). Right: the $p_T^{\Lambda}$ distributions of data (points) and
simulation (histogram), each normalized to unit area, and their ratio, the reweighting factor
$w(p_T)$.}
\label{fig:ptrw}
\end{figure}

The $\Lambda$ decays isotropically in the simulation, so the simulated $\cos\theta$ distribution along
any axis is the decay-angle acceptance of the selection. It is far from flat. The decay releases only
$\approx100\,\MeV$ of momentum in the $\Lambda$ rest frame, so the proton, by its mass, takes most of
the $\Lambda$ momentum in the laboratory irrespective of the decay angle, while the share of the pion
depends on the angle. When the proton is emitted forward ($\cos\theta>0$), the pion recoils backward and its
laboratory momentum is minimal, so that it falls most often below the threshold of $0.35\,\GeV$ that
the central $V^0$ reconstruction imposes on the transverse momentum of each daughter track and the
candidate is lost. The acceptance therefore disfavors forward protons, and the unpolarized simulation has
$\langle\cos\theta\rangle<0$. The loss is largest for soft $\Lambda$, whose pions are the
softest: in the control region defined below, $\langle\cos\theta\rangle$ varies steeply with the
$\Lambda$ transverse momentum (figure~\ref{fig:ptrw}, left), from a strongly negative value for the
softest candidates to a slightly positive one for the hardest candidates, among which the share of
random track pairs grows. For a random pair of tracks, which in most cases are two pions, the track with the
larger laboratory momentum is the one emitted forward in the rest frame of the pair, so the assignment
of the proton mass to the harder track places the proton forward, at $\cos\theta>0$.

The $\Lambda$ $p_T$ spectrum is, in addition, markedly softer in data than in simulation
(figure~\ref{fig:ptrw}, right). We find that candidates whose line of flight misses the primary vertex
by more than 0.5~cm along the beam, referred to below as \emph{the far class}, account for
\PtrwFarSoftSharePct\% of the data excess over the simulation below $3.5\,\GeV$. These are mostly $\Lambda$
baryons from pileup interactions. In the simulation, \DzFarNonGenPct\% of them are not matched to a
generator-level $\Lambda$, a record that pileup particles lack, and their mass spectrum peaks at
$m_\Lambda$. They are charge-symmetric
($\DzFarRSWSData$ right-minus-wrong-sign in data), and the simulation under-produces them, \DzFarSim\
simulated compared with \DzFarData\ observed in the signal window, counting both pairings of the baryon
number with the lepton charge, whereas the candidates
within 0.2~cm of the primary vertex agree (\DzNearSim\ simulated, \DzNearData\ observed). The candidates that
point to the primary vertex are themselves softer in data than in simulation. For these, a deficit of
soft strange baryons in the generator would be consistent with earlier LHC measurements of
strange-hadron production~\cite{cms-strangeprod,alice-strangeness2017}, but the present data do not
separate such a deficit from a mismodeling of the reconstruction efficiency for soft, displaced $V^0$
candidates. The reweighting described below corrects the spectrum irrespective of its origin.

The simulation is reweighted by $w(p_T)=f_{\mathrm{data}}(p_T)/f_{\mathrm{sim}}(p_T)$, the ratio of
the normalized $p_T$ distributions of data and simulation in a control region, in the bins of
figure~\ref{fig:ptrw}. The control region consists of the $\Lambda$ candidates of the signal-region
events that pass all requirements of section~\ref{sec:v0} in any of the three jets of the top triplet.
The charm-tag requirement of section~\ref{sec:hadtop}, which identifies the strange jet, is not
applied. The control region contains $N=\PtrwNctl$ data candidates, \PtrwBjetSharePct\% in
the $b$ jets and \PtrwWjetSharePct\% in the two $W$-candidate jets. The strange-jet sample in which $P$
is measured makes up only \PtrwSjetSharePct\% of it. The weights are therefore determined primarily by
candidates outside the measured sample, and the region is nearly unpolarized along the longitudinal
axis. Its polarized components are these few percent of strange-jet candidates and the $\Lambda$ baryons
from charmed-hadron decays in the $b$ and $c$ jets, \PtrwCharmHadronPct\% of the simulated candidates,
whose decay polarization the simulation does not reproduce; both enter the closure that will be presented in
section~\ref{sec:v0checks}.

Within each $\Lambda$-baryon $p_T$ bin, $\langle\cos\theta\rangle$ of the data
agrees with that of the unpolarized simulation ($\chi^2=\PtrwCosChiSq$ for the nine bins of
figure~\ref{fig:ptrw}, left). Because the acceptance depends on $p_T$ and the $p_T$ spectra differ,
the inclusive means nevertheless differ: in the control region the data give
$\langle\cos\theta\rangle=\CtlMeanData$, and the simulation $\CtlMeanSimRaw$ before the reweighting and
$\CtlMeanSimRW$ after it. Without the reweighting, the difference would be read as a polarization.

\begin{figure}[tp]
\centering
\includegraphics[width=\textwidth]{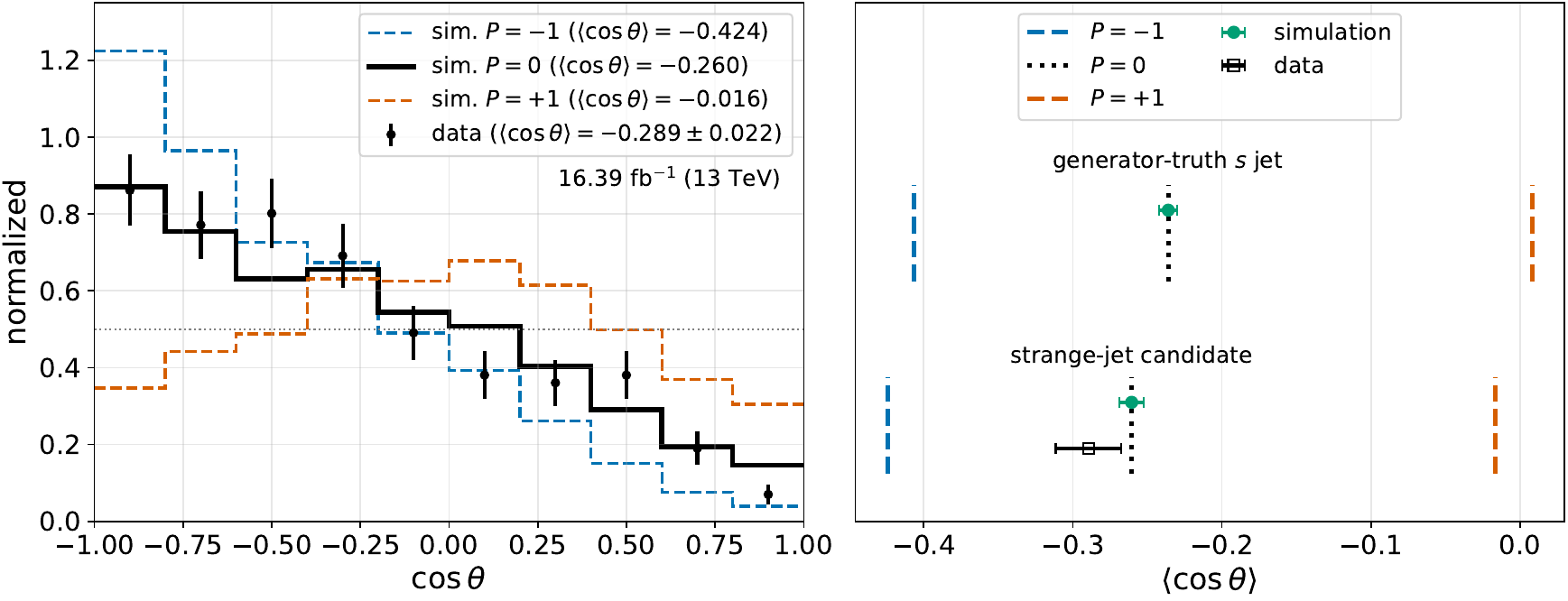}
\caption{$\Lambda$ polarimetry along the longitudinal axis in the strange jet of $W^+\to c\bar s$. Left: the $\Lambda$ decay angle
$\cos\theta$ in the unpolarized simulation (semileptonic and dileptonic \ttbar and single
top), with the $\Lambda$ \pt reweighting of the polarization fit (solid), together with
its reweighting by $1+\alpha P\cos\theta$ to $P=-1$ and $+1$ (dashed) and the collision data
(points). Since only a fraction of the
candidates carry the primary strange quark, their physical polarization shifts this inclusive
sample by far less than the templates shown. Right:
$\langle\cos\theta\rangle$ for two samples of simulated candidates, those whose jet is a strange jet
of $W\to c\bar s$ at the generator level (generator-truth $s$ jet) and those in the $W$ jet
recoiling against the charm-tagged jet (strange-jet candidate). For each sample
the simulated value (filled) is shown with the $P=0$ and $P=\pm1$ expectations of its templates
(lines), and for the strange-jet candidate also the data (open).}
\label{fig:sjetpol}
\end{figure}

Polarization templates are obtained from the unpolarized simulation by weighting each simulated event by
$1+\alpha P\cos\theta$, which injects a polarization $P$ on top of the same acceptance.
Figure~\ref{fig:sjetpol} (left) shows the $\cos\theta$ distribution of the $\Lambda$ candidates in the
strange jets: the unpolarized simulation, the $P=\pm1$ templates and the collision data. The templates
illustrate the response of $\langle\cos\theta\rangle$: $P=\pm1$ moves the mean by \TemplateShiftMinus\
and \TemplateShiftPlus, asymmetrically because the unpolarized acceptance is itself asymmetric, where a
flat acceptance would give $\alpha P/3\approx\pm\CosFullP$ for both.\footnote{A reweighting by
$1+\alpha P\cos\theta$ moves the mean from $\mu$ to
$(\mu+\alpha P\langle\cos^2\theta\rangle)/(1+\alpha P\mu)$, with $\mu$ and
$\langle\cos^2\theta\rangle$ those of the unpolarized acceptance.}

\subsection{Estimator, signal content of the sample, and the observable of interest}
\label{sec:v0obs}

Each polarization component is extracted with an unbinned maximum-likelihood fit of
$1+\alpha P\cos\theta$ to the decay angles of the candidates, in which the acceptance is taken into
account through the simulated mean $\langle\cos\theta\rangle$ of the
unpolarized template. Since the probability density of $\cos\theta$ is the unpolarized
acceptance multiplied by $1+\alpha P\cos\theta$ and normalized, the likelihood is
\begin{equation}
\ln L(P)=\sum_i\ln(1+\alpha P\cos\theta_i)-N\ln(1+\alpha P\langle\cos\theta\rangle_{\mathrm{sim}})
\label{eq:likelihood}
\end{equation}
up to $P$-independent terms, irrespective of the shape of the acceptance, with $N$ the number of
data candidates.

Such a fit returns the polarization of the whole selected sample. Only a minority of its candidates
can inherit the strange-quark polarization: the $\Lambda$ baryons formed from the strange quark of the
$W$ decay, directly or through the decay of a heavier hyperon. The baryon number
of such a $\Lambda$ follows that of the quark, which is fixed by the charge
of the parent top quark and hence by the lepton charge, so it is \emph{right-sign} by construction,
i.e.\ it satisfies the lepton-charge requirement of section~\ref{sec:v0}. The other candidates are
to first approximation charge-symmetric and unpolarized: they populate the
right-sign selection and the opposite, \emph{wrong-sign} pairing (a $\bar\Lambda$ with an $\ell^+$ or
a $\Lambda$ with an $\ell^-$) equally. Writing $S$ for the former and $B$ for the latter,
$N_{\mathrm{RS}}=S+B/2$ and $N_{\mathrm{WS}}=B/2$, so the charge-correlated fraction
$g\equiv(N_{\mathrm{RS}}-N_{\mathrm{WS}})/(N_{\mathrm{RS}}+N_{\mathrm{WS}})=S/(S+B)$ is read directly
off the yields, and the charge-correlated purity of the right-sign sample is
$f=S/N_{\mathrm{RS}}=2g/(1+g)$. Inclusively in $z$, $g=\gInclData$ in the data and $\gInclSim$ in
simulation, i.e.\ $f\approx\PurityInclSimPct\%$. The purity is low because the inclusive selection is
dominated by soft fragmentation $\Lambda$ (mean momentum fraction $\langle z\rangle=\ZmeanSjetData$ in
the data). Because the unpolarized component enters data and simulation alike and shares the decay-angle
acceptance, the fit returns the diluted polarization $P_{\mathrm{fit}}=f\,P(\Lambda)$, where $P(\Lambda)$ is the polarization of the
charge-correlated $\Lambda$. The purity $f$ is a net quantity rather than the fraction of $\Lambda$
baryons that carry the primary quark: the $\Lambda$ of charm jets are anti-correlated with the lepton charge,
the charm side of the $W$ decay being the antiparticle partner of the strange side, and partly cancel
the strange excess. Restricting the simulation to truth-strange jets, where no cancellation occurs,
raises the inclusive purity to $\PurityInclStrangeSimPct\%$.

The purity rises with the momentum fraction, high-$z$ $\Lambda$ baryons being the more likely to carry the
primary strange quark (figure~\ref{fig:zsplit}, top right). In simulation $g$ grows to $\gZthreeSim$
for $z>0.3$, so that $f$ climbs to $\PurityZthreeSimPct\%$, and to $\PurityZthreeStrangeSimPct\%$ within
truth-strange jets; the data give $g=\gZthreeData$ there, consistent with the simulation.

The interpretation of $P_{\mathrm{fit}}$ depends on what is counted as signal. For a class $X$ of
candidates, $P_{\mathrm{fit}}=D_X\,P_X$, with $D_X$ the size of the class relative to the fitted sample
and $P_X$ its polarization, provided that the other classes are unpolarized and share the decay-angle
acceptance. Two conventions are used below, at $z>0.3$. The charge-correlated purity, $D_f=f$, is
$\LamDilF$ in the simulation and $\LamDilFData$ in the data. It is the only ratio measurable in data,
and it understates the fraction of candidates formed from the strange quark of the $W$ decay: the
anti-correlated $\Lambda$ baryons of charm and bottom jets are subtracted from it, whereas the $\Lambda$
baryons formed from sea $s\bar s$ pairs are charge-symmetric and cancel in it. The ratio
$P_{\mathrm{fit}}/D_f$ therefore overstates the magnitude of the polarization of the charge-correlated
$\Lambda$ baryons. The ensemble ratio,
$D_{\mathrm{ens}}=\LamDilEns$ in the simulation, counts every genuine $\Lambda$ baryon of a
strange-flavored jet, of either baryon number, per right-sign candidate. This is the ensemble of the LEP
analyses, in which the dilution by the $\Lambda$ baryons
formed from secondary strange quarks is part of the predicted polarization, and the band quoted in
section~\ref{sec:intro} refers to it. The wrong-sign members of the ensemble are formed from secondary
strange quarks and are unpolarized. They are not among the fitted candidates, but they reduce the
magnitude of the ensemble polarization and are therefore counted in $D_{\mathrm{ens}}$. That the other
classes are unpolarized is an approximation: $\Lambda$ baryons that carry the primary quark of a
light-quark jet entering the sample can carry a polarization transferred
from that quark, which is expected to be small and, for the $u$ and $d$ quarks, either vanishing or
of the sign opposite to that of the strange quark, depending on the
model~\cite{burkardt-jaffe,jaffe-lambdafrag,kotzinian-bravar,boros-liang}; $\Lambda$ baryons from
$\Lambda_c$ decays are polarized but populate mostly the wrong-sign sample, as will be discussed in
section~\ref{sec:v0checks}; and
the $\Lambda$ baryons formed from sea $s\bar s$ pairs and the combinatorial and pileup candidates carry none.

The observable of interest of this analysis is the longitudinal polarization in the $z>0.3$ region,
where it is expected to be sizable. The band of section~\ref{sec:intro},
$P(\Lambda)\approx-\PLamBandLo$ to $-\PLamBandHi$ at $z>0.3$, is the prediction for the ensemble
defined above with the polarization $P(s)\approx-\PsZBorn$ of the strange quarks of $Z$ decays. For
the fully polarized strange quark of the $W$ decay the same transfer gives
$P(\Lambda)\approx-\PLamBandWLo$ to $-\PLamBandWHi$.\footnote{The band is scaled by the ratio
of the Born-level quark polarizations, $1/\PsZBorn$; the $\mathcal{O}(\alpha_s)$ corrections of
footnotes~\ref{fn:psz} and~\ref{fn:psw} change the ratio by less than 2\%.} With
$D_{\mathrm{ens}}=\LamDilEns$, such a polarization corresponds to $P_{\mathrm{fit}}$ between
$\ExpPfitZthreeLo$ and $\ExpPfitZthreeHi$, i.e.\ to a shift of $\langle\cos\theta\rangle$ by about
$\ExpCosShiftZthree$.

\subsection{Results}
\label{sec:v0res}

Table~\ref{tab:sjetpolP} lists the acceptance-corrected polarization along the three axes for the
full sample, for $z>0.2$, and in the nominal $z>0.3$ signal region, with the wrong-sign control of
section~\ref{sec:v0checks} beside each value.
The observable of interest is
\begin{equation}
P_\parallel(z>0.3)=\PparZthree\ \mathrm{(stat.)}\pm\PtrwSyst\ \mathrm{(syst.)},
\label{eq:pzthree}
\end{equation}
consistent with zero and with the expected $\ExpPfitZthreeLo$ to
$\ExpPfitZthreeHi$; the systematic uncertainty is that of the $\Lambda$ \pt reweighting
(section~\ref{sec:v0checks}). The
looser $z>0.2$ region gives $\PparZtwo$, and the full sample gives $\PparFull$. The $\Lambda$ and
$\bar\Lambda$ halves of the full sample, each fitted with its own simulated template, give
$P_\parallel=\PparLam$ and $\PparLbar$, a difference of $\PparLamLbarDiff$.
A baryon-number-odd acceptance mismatch between data and simulation, such as a proton--antiproton
material-interaction asymmetry, would appear here. Both transverse components are consistent with
zero, as expected.

\begin{table}[tp]
\centering
\begin{tabular}{lrrrr}
\toprule
 & & & \multicolumn{2}{c}{wrong-sign control} \\
\cmidrule(lr){4-5}
Component & $N_{\mathrm{data}}$ & \multicolumn{1}{c}{$P$} & \multicolumn{1}{c}{$P_{\mathrm{WS}}$} & \multicolumn{1}{c}{$P-P_{\mathrm{WS}}$} \\
\midrule
\Sjetpolpbody
\bottomrule
\end{tabular}
\caption{Acceptance-corrected $\Lambda$ polarization in the strange jet of $W^+\to c\bar s$ along
the three decay axes with no $z$ cut, with $z>0.2$, and with the nominal $z>0.3$, from the unbinned
maximum-likelihood fit of section~\ref{sec:v0obs}. $N_{\mathrm{data}}$ is the number of $\Lambda$
baryons of the strange-jet sample entering each fit. The last two columns give the same fit on the wrong-sign control
sample, with the template of its own simulated sample, $P_{\mathrm{WS}}$, and the wrong-sign-subtracted
polarization, $P-P_{\mathrm{WS}}$, with the two independent uncertainties added in quadrature
(section~\ref{sec:v0checks}). Uncertainties are statistical only.}
\label{tab:sjetpolP}
\end{table}

\subsection{Cross-checks and systematic uncertainty}
\label{sec:v0checks}

\emph{Wrong-sign acceptance control.} The wrong-sign sample of section~\ref{sec:v0obs} shares the
strange-jet selection, the $V^0$ reconstruction and the combinatorial background of the right-sign sample but
receives no polarization from the $W\to c\bar s$ decay. Polarized admixtures arise from two sources. The
$\Lambda$ baryons of charmed-hadron decays enter through charm jets misassigned as the strange-jet
candidate, and their share is bounded by the $\WSCjetPct\%$ of the wrong-sign candidates that lie in
truth-charm jets in simulation. The $\Lambda$ baryons formed from the primary quark of a $u$-quark jet
are polarized as well, whereas those of a $d$-quark jet are right-sign, like those of the strange quark.
Both admixtures are negligible at the present precision, the second because the polarization transferred
from a $u$ quark is expected to be small (section~\ref{sec:v0obs}). Table~\ref{tab:sjetpolP} lists the
polarization of the wrong-sign sample and the wrong-sign-subtracted polarization beside the right-sign
values for every axis and $z$ selection. Passed through the same acceptance-corrected fit, with the
template of its own simulated sample, the full wrong-sign sample returns $P_\parallel=\PwsLong$, with
the in-plane and normal components $\PwsInpl$ and $\PwsNorm$, consistent with zero, as expected for a
polarization-free sample. An acceptance or background effect common to the two pairings would appear in
both samples. The wrong-sign sample shows none, with the same statistical precision as the right-sign
value $\PparFull$.

\emph{Reweighting closure.} In the control region the fit moves from $P_\parallel=\PtrwCtlRaw$
before the $\Lambda$ $p_T$ reweighting to $\PtrwCtlCorr$ after it. The residual, consistent with zero
within its statistical uncertainty, bounds the combined effect of the mismodeling that a function of
$p_T$ alone does not capture and of the polarization that the $\Lambda$ baryons from weak decays of
charmed hadrons in the $b$ and $c$ jets of the region have in the data but not in the simulation.
Derived separately in the $b$ jet and in the $W$-candidate jets of the region, the weights give
$\chi^2=\PtrwClassChiSq$ for \PtrwClassNdf\ degrees of freedom in their bin-by-bin comparison, and the
fitted polarization changes by less than \PtrwClassShiftMax\ with either set. The residual after the
reweighting is $\PtrwCtlCorrB$ in the $b$ jet and $\PtrwCtlCorrW$ in the $W$-candidate jets outside the
strange-jet sample. The larger of the two, that of the
$W$-candidate jets, is assigned in magnitude as a systematic uncertainty of $\PtrwSyst$ on every
polarization quoted in section~\ref{sec:v0data}, smaller than the statistical uncertainty throughout.
In the $z>0.3$ signal region the reweighting changes $P_\parallel$ from $\PparZthreeRaw$ to
$\PparZthree$. The shift here is very small because the high-$z$ requirement already removes the soft
$\Lambda$ where the acceptance is steepest. On the full sample it moves $P_\parallel$ from $\PparFullRaw$ to $\PparFull$.
Fitting $P_\parallel$ separately in each $\Lambda$ $p_T$ bin and combining the bins, so that no
reweighting model enters, gives $\PtrwDiffZthree$ for $z>0.3$ and $\PtrwDiffSR$ on the full sample,
consistent with the reweighted fits.

\emph{Pointing requirement.} Requiring every $\Lambda$ candidate to point to the primary vertex within
$|d_z|<0.5$~cm, which removes the far class of section~\ref{sec:v0acc}, would cost \DzCutLossZthreePct\% of
the data candidates at $z>0.3$ and change $P_\parallel(z>0.3)$ from $\PparZthree$ to $\PparZthreeDz$, a
shift of $\DzCutShiftZthree$, or \DzCutShiftZthreeSig\ standard deviations of the difference between the nested
samples. The uncertainty of the difference is the spread over bootstrap replicas of the data and of the
simulation, with both nested samples formed from each replica.
In the simulation the removed candidates are about as charge-correlated as the retained ones ($g=\DzZthreeCutG$
and $\DzZthreeKeepG$). The correlation of the removed candidates comes
entirely from the \DzZthreeCutGenPct\% of them that are genuine $\Lambda$ baryons at large flight distance.
The others are uncorrelated. The requirement would thus raise the charge-correlated
purity of the simulated sample only from $\PurityZthreeSimPct\%$ to $\DzKeepPurityZthreePct\%$, and the data
cannot test the composition of the removed candidates at the present statistics. It is not applied in our analysis.

\emph{Estimator validation.} The total statistical uncertainty of the fit contains, besides the data-statistical
term, the statistical uncertainty of the simulated template, which enters through
$\langle\cos\theta\rangle_{\mathrm{sim}}$: after the event weights and the $p_T$ reweighting the
simulated sample has an effective size of only about \NmcEffRatio\ times that of the data, so the
template term is about a third of the data term; it is propagated analytically and added in quadrature.
A separate test validates the estimator itself, with
pseudo-experiments drawn from the unpolarized simulation and fitted with the same simulated template, so
that only the data-statistical term enters. The closure is unbiased, with a pull width of $\UnbPull$,
and an injected $P_\parallel=-0.20$ is recovered as $\UnbInjOut$.
The pseudo-experiments do not test the smearing of the polarimeter axis by the reconstructed-$W$ boost,
because the injection, like the fit, defines the polarization along the reconstructed axis. This
smearing dilutes a polarization along the true axis by about $\SmearDilPct\%$, a multiplicative
correction negligible at the present precision.

\subsection{Projections}
\label{sec:v0reach}

The present work demonstrates the analysis chain on part of the 2016 data. A measurement will require
substantially more integrated luminosity, together with a data-driven estimate of the backgrounds
and the corresponding systematic uncertainties. The reach we will quote here follows from the statistical
scaling alone, with the simulated sample assumed to grow with the data so that the data-statistical
and the template terms of the uncertainty scale alike. Table~\ref{tab:lamreach} scales the present
uncertainty on $P_{\mathrm{fit}}$ at $z>0.3$, $\sigma_{P_{\mathrm{fit}}}=\dPUnbTotZthree$, with the number of $\ttbar$ events
expected for Run~2, for Run~2 and~3 combined and for the HL-LHC.

\begin{table}[tp]
\centering
\renewcommand{\arraystretch}{1.2}
\begin{tabular}{lccc}
\toprule
 & Run~2 & Run~2\,+\,3 & HL-LHC \\
 & $140\,\mathrm{fb}^{-1}$ & $450\,\mathrm{fb}^{-1}$ & $3000\,\mathrm{fb}^{-1}$ \\
\midrule
$\sigma_{P_{\mathrm{fit}}}$ (fitted sample)                                       & \LamReachRunTwo     & \LamReachRunThree     & \LamReachHL     \\
$\sigma_{P_{\mathrm{fit}}}/D_f$ \ ($D_f=\LamDilF$)                                & \LamReachRunTwoF    & \LamReachRunThreeF    & \LamReachHLF    \\
$\sigma_{P_{\mathrm{fit}}}/D_{\mathrm{ens}}$ \ ($D_{\mathrm{ens}}=\LamDilEns$)    & \LamReachRunTwoEns  & \LamReachRunThreeEns  & \LamReachHLEns  \\
\bottomrule
\end{tabular}
\caption{Projected statistical reach on the $\Lambda$ polarization, scaled
from the present $z>0.3$ uncertainty with the number of $\ttbar$ events, $\sigma_{\ttbar}L$, with the
NNLO+NNLL cross sections at 13, 13.6 and $14\,\TeVns$~\cite{toppp} for Run~2 ($140\,\mathrm{fb}^{-1}$),
Run~3 ($310\,\mathrm{fb}^{-1}$) and the HL-LHC. The first row is the uncertainty
$\sigma_{P_{\mathrm{fit}}}$ on the fitted sample; the next two convert it to the
physics reach $\sigma_{P_{\mathrm{fit}}}/D$ under the two dilution conventions $D_f$ (the wrong-sign--subtracted
excess, measurable in data) and $D_{\mathrm{ens}}$ (the genuine $\Lambda$ baryons of either baryon number in a
strange-flavored jet per right-sign candidate, the LEP convention), defined in section~\ref{sec:v0obs}. Columns give the projected integrated luminosities of full
Run~2, Run~2 and~3 combined, and the HL-LHC.}
\label{tab:lamreach}
\end{table}

Three improvements are available for a future measurement. First, part of the charm-jet
contamination of the strange-jet sample arises from the mis-pairing of jets in the $\chi^2$
reconstruction of the hadronic top quark. A kinematic likelihood over the full event would raise the
strange- and charm-jet purities at fixed efficiency. Second, the wrong-sign control would enter a joint
likelihood over the right-sign and wrong-sign samples, with the
signal polarization and a charge-symmetric acceptance asymmetry common to both samples as its
parameters. With the common asymmetry free, the joint fit is the subtraction of
section~\ref{sec:v0checks}, and with it fixed the wrong-sign sample adds nothing to the polarization. A
measurement would constrain the asymmetry by the simulation, the control region of
section~\ref{sec:v0acc} and the wrong-sign sample together, so that the wrong-sign sample tests the
acceptance without doubling the statistical uncertainty. Third, a longitudinal pointing requirement, the analogue of the transverse
pointing angle of section~\ref{sec:v0}, which the central $V^0$ reconstruction does not apply, targets
the $\Lambda$ baryons from pileup interactions, the one component of the sample that the simulation
under-produces (section~\ref{sec:v0acc}). In the simulation, inclusive in $z$, it would remove
\DzCutLossNonGenFullPct\% of the candidates not matched to a generator-level $\Lambda$ at a cost of
\DzCutLossGenFullPct\% of the genuine ones, and it would reduce the $\Lambda$ $p_T$ reweighting, whose
residual is the systematic uncertainty of section~\ref{sec:v0checks}. The weights $w(p_T)$ of
section~\ref{sec:v0acc} would range from \PtrwWmaxDz\ in the softest $p_T$ bin to \PtrwWminDz\ in
the hardest instead of from \PtrwWmax\ to \PtrwWmin. The requirement is not applied here, where it would change
the $z>0.3$ purity little (section~\ref{sec:v0checks}), but it is the selection we recommend for a
measurement, once its effect on the decay-angle acceptance has been validated in data. The inefficiency of the requirement for genuine
$\Lambda$ baryons is concentrated at large flight distances, where the daughter tracks start beyond the pixel
layers, and the softer of the two tracks, the pion, has its momentum set by the decay angle, so the
inefficiency could depend on $\cos\theta$. The wrong-sign sample and the control region of
section~\ref{sec:v0acc}, fitted with and without the requirement, would test this.

The projection to the HL-LHC assumes the present detector. The Phase-2 upgrades of the CMS tracking
system~\cite{cms-phase2-tracker,cms-mtd} are of direct relevance to this analysis, which relies on
displaced vertices of soft tracks inside jets. The new tracker will extend the coverage from
$|\eta|<2.5$ to $|\eta|<4$, with a substantially reduced material budget, and replace the present
pixels of $100\times150\,\mu\mathrm{m}^2$ by cells of $25\times100\,\mu\mathrm{m}^2$. The
reconstruction efficiency and the impact-parameter resolution of soft tracks, both limited by
multiple scattering, will improve accordingly, and the rate of photon conversions and nuclear
interactions that mimic displaced $V^0$ vertices will decrease. The decay-angle acceptance, which is
shaped by the loss of the soft pion in decays with a forward proton (section~\ref{sec:v0acc}), will
become flatter, and the $p_T$ reweighting correspondingly smaller. The MIP timing detector will
provide a time measurement with a resolution of 30--40 ps per track, which will associate each
track, and through its daughters each $V^0$ candidate, with its interaction vertex. This
association will remove the pileup component that accounts for
the data excess over the simulation at low $p_T$ (section~\ref{sec:v0acc}). The $b$ and charm tagging,
which build on the same track impact parameters and secondary vertices, would improve
correspondingly~\cite{cms-phase2-tracker,cms-mtd}, raising the efficiency of the signal-region selection
and the purity of the strange-jet sample at a given efficiency. These gains are not
included in table~\ref{tab:lamreach}.

\section{Conclusions}
\label{sec:conclusions}

A first realization of strange-quark polarimetry on the polarized strange quarks of the hadronic $W$
decay in semileptonic \ttbar events has been presented on the public CMS data of 2016. The analysis
chain comprises the reconstruction of the hadronic top quark, the identification of the
strange jet as the $W$-decay jet recoiling against the charm-tagged jet, the selection of
$\Lambda\to p\pi^-$ candidates with the baryon number expected from the lepton charge, and a fit of the
proton decay angle, measured relative to the $\Lambda$ direction in the $W$ rest frame, to templates
obtained from the simulation.

The observable of interest is the longitudinal polarization of the $\Lambda$ in the region $z>0.3$ of
the momentum fraction carried by the $\Lambda$ within the jet, where the candidates that contain the
original strange quark are most abundant. It is found to be
$P_\parallel=\PparZthree\ \mathrm{(stat.)}\pm\PtrwSyst\ \mathrm{(syst.)}$, consistent both with zero
and with the value expected from the polarization transfer inferred from the LEP data. The two
transverse components are consistent with zero, as expected. The present precision is insufficient to
test the polarization transfer. The contribution of the analysis is the validated chain and the
findings on which a measurement with larger datasets can be built.

Two findings are relevant to the design of such a measurement. The first concerns the composition of
the sample. Most $\Lambda$ candidates are soft fragmentation baryons unrelated to the polarized quark,
and the fraction that contains the strange quark rises with the momentum fraction $z$. The correlation
between the baryon number and the lepton charge measures this fraction in data, and the wrong-sign
sample, the candidates with the baryon number opposite to the expected one, provides a control of the
acceptance in data. It shares the reconstruction and the background of the signal sample but carries
no polarization from the $W$ decay. The second concerns the acceptance. The decay-angle acceptance
depends steeply on the $\Lambda$ transverse momentum, and the $\Lambda$ spectrum is softer in data
than in simulation. Unless the simulation is reweighted to the spectrum of the data, the mismatch
mimics a polarization. The residual after the reweighting, the one systematic uncertainty evaluated
here, is smaller than the statistical one. The simulation under-produces $\Lambda$ baryons from pileup
interactions, and these account for most of the excess of soft $\Lambda$ baryons in the data. Such
candidates are identified by the distance of their trajectory from the primary vertex along the beam.
They are equally frequent for both baryon numbers and therefore dilute the polarization without
biasing it. A pointing requirement would remove most of them, at the cost of a few percent of the
genuine $\Lambda$ baryons that decay far from the beam line, where the extrapolation of their line of
flight to the beam is least precise. It is not applied here, since in the simulation the purity of the
$z>0.3$ sample would rise only marginally and the data cannot test the composition of the removed
candidates. For a measurement we recommend the pointing requirement as the handle on the one
component of the sample that the simulation under-produces, once its effect on the decay-angle
acceptance has been validated in data.

Scaled with the number of \ttbar events, the statistical uncertainty on the $\Lambda$ polarization in
strange jets reaches about $\LamReachRunTwoEns$ with the full Run-2 data of one experiment and
$\LamReachHLEns$ at the HL-LHC, compared with the predicted $P(\Lambda)\approx-\PLamBandWLo$ to
$-\PLamBandWHi$ (section~\ref{sec:v0obs}). The latter uncertainty is of the size
of that of the LEP measurements. The two quantities differ: the LEP value is the polarization of the
inclusive $\Lambda$ sample, averaged over the quark flavors, whereas the value here would be that of the
$\Lambda$ baryons of strange jets alone. A kinematic likelihood for
the jet assignment and a joint fit of the right-sign and wrong-sign samples would raise the purity and the
statistical efficiency. The tracking and timing detectors of the HL-LHC upgrade would flatten the acceptance,
remove the pileup component and sharpen the flavor tagging. A measurement with the
precision expected at the HL-LHC would require a data-driven estimate of the backgrounds and of the systematic uncertainties of
the acceptance correction, which the present analysis has identified but not quantified at that level.

The LEP measurements averaged over the quark flavors, and their interpretation in terms of the
strange quark relied on models of the flavor composition. The strange jets of the hadronic $W$ decay,
identified through their charm partner, isolate the strange quark. The same events provide, with the
same reconstruction, $\Lambda$ baryons from $b$ jets, from charm-tagged jets and from the light-quark
jets of $W\to u\bar d$ decays, each with a known quark polarization, so that the spin transfer can be
decomposed by flavor in the data. The full Run-2 samples of ATLAS and CMS offer the opportunity for
such a decomposition.

\section*{Acknowledgements}
We thank the CMS Collaboration for making their data publicly available.
The analysis software, statistical tools, figures, and text of this manuscript were
prepared with extensive use of Anthropic's Claude models (Opus 4.8, 5, and~5.5 and
Fable 5 and~5.1), through the Claude~Code coding assistant for the
data analysis and the Claude desktop application for the planning and review of the work.
All results and text were thoroughly scrutinized by the
author, who takes full responsibility for the content of this paper.

\paragraph{Funding information}
Y.K. was supported in part by the Israel Science Foundation (Grant No.~1666/22).

\renewcommand{\bibsection}{\section*{\phantomsection\tocbibname}\addcontentsline{toc}{section}{\tocbibname}}
\bibliography{CMS-ttbar-Lambda}

\end{document}

%% file: note_numbers.tex
\def\LumiDet{16.39}
\def\AlphaL{0.755}
\def\CosFullP{0.25}
\def\KsVetoCtr{496.7}
\def\KsVetoOffset{0.9}
\def\PLamBandLo{0.42}
\def\PLamBandHi{0.55}
\def\PsZBorn{0.94}
\def\PLamBandWLo{0.45}
\def\PLamBandWHi{0.59}
\def\AlphaLEP{0.642}
\def\LMCfb{175.7}
\def\NMCmini{63.96}
\def\DlNev{14.06}
\def\DlLeff{159.8}
\def\StTWtNev{2.49}
\def\StTWtLeff{69.5}
\def\StTWatNev{2.55}
\def\StTWatLeff{71.2}
\def\StTctNev{8.46}
\def\StTctLeff{62.2}
\def\StTcatNev{5.14}
\def\StTcatLeff{63.4}
\def\TemplateShiftMinus{$-0.16$}
\def\TemplateShiftPlus{$+0.24$}
\def\PurMed{37}
\def\Sjetpolpbody{Longitudinal               &  499 & $-0.16\pm0.13$ & $+0.00\pm0.13$ & $-0.16\pm0.18$ \\
\quad $z>0.2$              &   95 & $-0.26\pm0.28$ & $+0.07\pm0.28$ & $-0.33\pm0.40$ \\
\quad $z>0.3$ (nominal)    &   37 & $+0.04\pm0.45$ & $+0.34\pm0.39$ & $-0.30\pm0.60$ \\
\midrule
Normal to plane            &  499 & $+0.02\pm0.11$ & $-0.06\pm0.11$ & $+0.08\pm0.15$ \\
\quad $z>0.2$              &   95 & $-0.05\pm0.24$ & $-0.06\pm0.28$ & $+0.01\pm0.37$ \\
\quad $z>0.3$ (nominal)    &   37 & $+0.10\pm0.37$ & $-0.34\pm0.46$ & $+0.43\pm0.59$ \\
\midrule
In-plane                   &  499 & $-0.11\pm0.11$ & $+0.00\pm0.11$ & $-0.11\pm0.16$ \\
\quad $z>0.2$              &   95 & $-0.07\pm0.23$ & $+0.16\pm0.29$ & $-0.22\pm0.37$ \\
\quad $z>0.3$ (nominal)    &   37 & $+0.22\pm0.36$ & $-0.10\pm0.51$ & $+0.33\pm0.62$ \\}
\def\PparFull{-0.16\pm0.13}
\def\PparZtwo{-0.26\pm0.28}
\def\PparZthree{+0.04\pm0.45}
\def\PparLam{-0.31\pm0.19}
\def\PparLbar{-0.01\pm0.18}
\def\PparLamLbarDiff{-0.30\pm0.26}
\def\PwsLong{+0.00\pm0.13}
\def\PwsInpl{+0.00\pm0.11}
\def\PwsNorm{-0.06\pm0.11}
\def\WSCjetPct{10}
\def\gInclData{+0.015\pm0.032}
\def\gInclSim{+0.031\pm0.012}
\def\gZthreeData{+0.16\pm0.12}
\def\LamDilFData{0.27\pm0.18}
\def\gZthreeSim{+0.28\pm0.04}
\def\PurityInclSimPct{6}
\def\PurityInclStrangeSimPct{15}
\def\PurityZthreeSimPct{43}
\def\PurityZthreeStrangeSimPct{79}
\def\ZmeanSjetData{0.129}
\def\CeffMed{0.34}
\def\LmisCm{0.054}
\def\BmisCm{0.15}
\def\NmcEffRatio{7.8}
\def\UnbPull{1.00}
\def\UnbInjOut{-0.20}
\def\dPUnbTotZthree{0.45}
\def\LamReachRunTwo{0.16}
\def\LamReachRunThree{0.08}
\def\LamReachHL{0.031}
\def\LamDilF{0.43}
\def\LamDilEns{0.62}
\def\LamReachRunTwoF{0.37}
\def\LamReachHLF{0.072}
\def\LamReachRunTwoEns{0.26}
\def\LamReachHLEns{0.050}
\def\LamReachRunThreeEns{0.13}
\def\LamReachRunThreeF{0.19}
\def\ExpPfitZthreeLo{-0.28}
\def\ExpPfitZthreeHi{-0.37}
\def\ExpCosShiftZthree{-0.07}
\def\ZfullMeanZthree{0.4}
\def\ZfullFracAbove{97}
\def\PtrwNctl{11\,084}
\def\PtrwBjetSharePct{44.7}
\def\PtrwWjetSharePct{55.3}
\def\PtrwSjetSharePct{4.5}
\def\PtrwCtlRaw{-0.189\pm0.026}
\def\PtrwCtlCorr{-0.014\pm0.026}
\def\CtlMeanData{-0.240\pm0.005}
\def\CtlMeanSimRaw{-0.201\pm0.002}
\def\CtlMeanSimRW{-0.238\pm0.002}
\def\PparFullRaw{-0.30\pm0.13}
\def\PparZthreeRaw{+0.03\pm0.45}
\def\PtrwDiffSR{-0.10\pm0.14}
\def\PtrwDiffZthree{+0.34\pm0.50}
\def\PtrwClassChiSq{14.4}
\def\PtrwClassNdf{8}
\def\PtrwClassShiftMax{0.01}
\def\PtrwCtlCorrB{+0.05\pm0.04}
\def\PtrwCtlCorrW{-0.06\pm0.04}
\def\PtrwCosChiSq{7.5}
\def\PtrwSyst{0.06}
\def\SmearDilPct{5}
\def\Cutflowbodynano{1 isolated lepton               & 1\,829\,685 & 575\,767 & 563\,683 & 2\,969\,135 & 229\,378\,245 & $77.25$ \\
$\ge$4 jets                     & 1\,308\,806 & 236\,943 & 131\,413 & 1\,677\,162 & 5\,699\,562 & $3.40$ \\
$\ge$1 $b$ jet                  & 1\,181\,134 & 213\,367 & 110\,371 & 1\,504\,872 & 2\,479\,731 & $1.65$ \\
$\ge$2 $b$ jets                 & 647\,077 & 120\,358 & 44\,791 & 812\,226 & 965\,900 & $1.19$ \\
hadronic top ($\chi^2<10$)      & 476\,827 & 59\,895 & 24\,164 & 560\,886 & 625\,892 & $1.12$ \\}
\def\Cutflowbodymini{hadronic top ($\chi^2<10$)      & 477\,157 & 59\,942 & 24\,184 & 561\,283 & --- & 625\,844 & $1.12$ \\
single $c$ tag                  & 93\,402 & 8\,642 & 4\,042 & 106\,086 & --- & 106\,296 & $1.00$ \\
$\ge$1 $\Lambda/\bar\Lambda$ in $s$ jet candidate & 1\,554 & 130 & 66 & 1\,750 & 119 & 1\,813 & $1.04 \pm 0.03$ \\
$\Lambda$ mass window           & 858 & 78 & 33 & 969 & 95 & 1\,069 & $1.10 \pm 0.04$ \\
$K_S$ veto                      & 764 & 71 & 29 & 864 & 90 & 979 & $1.13 \pm 0.04$ \\
$\Lambda/\bar\Lambda$ matching the lepton & 401 & 37 & 14 & 452 & 90 & 499 & $1.10 \pm 0.05$ \\
\midrule
$z>0.2$                         & 95 & 6 & 4 & 105 & 39 & 95 & $0.90 \pm 0.10$ \\
$z>0.3$                         & 47 & 2 & 1 & 50 & 25 & 37 & $0.74 \pm 0.13$ \\}
\def\CFShareDevData{0.0077}
\def\CFShareDevSim{0.071}
\def\DzFarData{356}
\def\DzFarSim{259}
\def\DzNearData{554}
\def\DzNearSim{577}
\def\DzFarNonGenPct{93}
\def\DzFarRSWSData{-6\pm19}
\def\PparZthreeDz{-0.09\pm0.46}
\def\DzCutLossZthreePct{14}
\def\DzCutShiftZthree{-0.13\pm0.12}
\def\DzCutShiftZthreeSig{1.1}
\def\PtrwWmax{1.87}
\def\PtrwWmin{0.71}
\def\PtrwWmaxDz{1.25}
\def\PtrwWminDz{0.78}
\def\DzZthreeCutGenPct{28}
\def\DzZthreeCutG{+0.20\pm0.10}
\def\DzZthreeKeepG{+0.29\pm0.04}
\def\DzKeepPurityZthreePct{45}
\def\DzCutLossGenFullPct{3.9}
\def\DzCutLossNonGenFullPct{58}
\def\PtrwFarSoftSharePct{89}
\def\PtrwCharmHadronPct{2.4}
\def\SigW{10.4}
\def\SigTop{18.4}
\def\MjjMedSim{81}
\def\MjjbMedSim{171}
\def\MjjbMedData{171}
\def\ChiSqSim{76}
\def\ChiSqData{68}
\def\ChiSqSimFull{72}
\def\LxyMed{6.4}
\def\LamProtonFrac{99.9}
\def\LamMassRes{1.5}
\def\LamMassNsig{3.9}
\def\LamMassKeep{82}
\def\DataMCdev{12}